\documentclass[]{aa}  
\usepackage{color }
\usepackage{natbib}
\usepackage{amsmath}
\usepackage{amssymb}
\usepackage{graphicx}
\usepackage{txfonts}
\usepackage{comment}
\usepackage{arydshln}
\newcommand{\bl}[1]{\mbox{\boldmath$ #1 $}}
\begin{document}

\title{Distinguishing between different mechanisms of FU-Orionis-type luminosity outbursts}
\titlerunning{Distinguishing between FU-Orionis-type outbursts}

\author{Eduard I. Vorobyov\inst{1,2}, Vardan G. Elbakyan\inst{2},  Hauyu Baobab Liu\inst{3}, and Michihiro Takami\inst{3}}
\institute{ 
University of Vienna, Department of Astrophysics, Vienna, 1180, Austria
\email{eduard.vorobiev@univie.ac.at} 
\and
Research Institute of Physics, Southern Federal University, Rostov-on-Don, 344090 Russia 
\and
Institute of Astronomy and Astrophysics, Academia Sinica, 11F of Astronomy-Mathematics Building, No.1, Sec. 4, Roosevelt Rd, Taipei 10617, Taiwan, R.O.C.
}
\authorrunning{Vorobyov et al.}


 
  \abstract
   {}
   {Accretion and luminosity bursts triggered by three distinct mechanisms: the magnetorotational instability in the inner disk regions, clump infall in gravitationally fragmented disks and close encounters with an intruder star, were studied to determine the disk kinematic characteristics that can help to distinguish between these burst mechanisms.}
   {Numerical hydrodynamics simulations in the thin-disk limit were employed to model the bursts in disk environments that are expected for each  burst mechanism. }
   {We found that the circumstellar disks featuring accretion bursts can bear kinematic features that are distinct for different burst mechanisms, which can be useful when identifying the burst origin.
   The disks in the stellar encounter and clump-infall models are characterized by tens of per cent deviations from the Keplerian rotation, while the disks in the MRI models are  characterized only a few per cent deviation, which is mostly caused by the gravitational instability that fuels the MRI bursts. Velocity channel maps also show distinct kinks and wiggles, which are caused by gas disk flows that are peculiar to each considered burst mechanism. The deviations of velocity channels in the burst-hosting disks from a symmetric pattern typical of Keplerian disks are strongest for the clump-infall and collision models, and carry individual features that may be useful for the identification of the corresponding burst mechanism. The considered burst mechanisms produce a variety of light curves with the burst amplitudes varying in the $\Delta m =2.5-3.7$ limits, except for the clump-infall model where $\Delta m$ can reach 5.4, although the derived numbers may be affected by a small sample and boundary conditions.}
   {Burst triggering mechanisms are associated with distinct kinematic features in the burst-hosting disks that may be used for their identification. Further studies including a wider model parameter space and the construction of synthetic disk images in thermal dust and molecular line emission  are needed to constrain the origin mechanisms of FU Orionis bursts.}


   \maketitle

\section{Introduction}

Protostars grow in mass through accretion from a surrounding protostellar disk. The details of protostellar accretion are, however, not fully understood. One aspect that is currently under debate is the behavior of protostellar accretion with time. Evidence is growing that protostellar accretion is not constant or steadily declining, but is highly variable \citep[e.g,][]{2017ContrerasPena,2020ZhenVISTA}. FU-Orionis-type luminosity outbursts (FUors) that are characterized by orders of magnitude increase in luminosity  provide a prime example of accretion variability. A few confirmed dozens of such objects are known to date \citep{2014AudardAbraham,2018ConnelleyReipurth} and several new candidates are routinely discovered every year.

It is generally agreed that FUors are caused by a sudden increase in the mass accretion rate from the disk on the protostar. However, the mechanisms that trigger such an increase are uncertain. These mechanisms include the magnetorotational instability  (MRI) in the innermost disk regions prompted by a sudden increase in the ionization fraction \citep[e.g.,][]{2001Armitage,2009ZhuHartmannGammie,2014BaeHartmann,2020Kadam}, infall of gaseous clumps formed through disk gravitational fragmentation  \citep[e.g.,][]{2005VorobyovBasu,2011MachidaInutsuka,2015VorobyovBasu,2017MeyerVorobyov}, planet-disk interaction and mass-exchange \citep{2004LodatoClarke,2012NayakshinLodato}, and close encounter between a protoplanetary disk and an intruder star in young stellar clusters \citep[e.g.][]{2008Pfalzner,2010ForganRice}. Global simulations of clustered star formation also indicate that gravitationally unstable protostellar disks with high rates of mass infall from the surrounding environment can drive accretion bursts \citep{2018KuffmeierFrimann}.  \citet{2014AudardAbraham} provides a comprehensive review on these and other burst triggering mechanisms.

Distinguishing between different outburst triggers is not an easy task. FUors show a variety of light curve shapes \citep{1996HartmannKenyon,2018ConnelleyReipurth}  and this was suggested as evidence for different underlying trigger mechanisms \citep{2014AudardAbraham}. 
The underlying physical conditions in a protoplanetary disk and its immediate environment may be vastly different.  Numerical simulations indicate that the same burst mechanism can show a spread in burst amplitudes and  durations, further complicating the comparison \citep[see e.g.,][]{2015VorobyovBasu}. 
An alternative approach to discriminate between different burst models would be to search for global disk features that may be particular to a certain burst mechanism. These can be the signatures of disk gravitational instability and fragmentation \citep{2014DunhamVorobyov,2018Cieza, 2019MeyerVorobyov} or close companions that might have caused the outburst \citep{2012BeckAspin} or disk winds and magnetic field structure that may be particular for the MRI-active inner disk \citep{2020Zhu}. In the recent study, \citet{2019MacFarlane} demonstrated that the flux increase during the burst is more prominent in the infrared than millimeter wavelengths but this effect may depend on the disk configuration, which in turn is specific to the triggering mechanism.

The theoretical comparison of different burst mechanisms is complicated by the use of different numerical codes and techniques \citep{2014AudardAbraham}. Even when using the same numerical code, the analysis of individual light curves during the bursts of different origin may be complicated by inherent differences in the numerical grid setup and uncertainties in the free parameters of the models. Therefore, in this paper we focus on the global kinematic signatures of burst-hosting disks, which are less affected by these difficulties and can help to distinguish between different burst mechanisms.
For this purpose, we employ the numerical hydrodynamics code FEOSAD (Formation and Evolution of Stars And Disks)  to explore the global dynamics of protoplanetary disks featuring FUor-type outbursts driven by three distinct mechanisms: MRI, clump infall, and close stellar encounter.

It has recently been shown that molecular line observations can reveal the presence of gravitational instability \citep{2020Hall} and the location of forming protoplanets \citep{2019Pinte} in the disk. Using our model data on the gas velocities, we constructed the velocity channel maps (which can serve as idealized proxies for CO channel-map emission)  and compare them with the channel maps obtained for an idealized unperturbed Keplerian disk to determine any specific kinematic signatures that can be used to distinguish between different burst mechanisms.

The paper is organized as follows.
Sect.~\ref{FEOSAD} presents the numerical model and discusses the considered burst mechanism.  Sect.~\ref{Kinematics} presents and compares the kinematic signatures of considered bursts.  Sect.~\ref{Caveats} provides the model caveats. Main conclusions are summarized in Sect.~\ref{Conclude}.

\section{Model description and considered burst mechanisms}
\label{FEOSAD}

We use numerical hydrodynamics simulations in the thin-disk limit to explore accretion bursts in young stellar systems. Three burst mechanisms were considered: triggering of the MRI in the innermost disk regions, infall of gaseous clumps in a gravitationally unstable disk, and encounter between a protoplanetary disk and an intruder (sub)-solar-mass star.
These models are referred to hereafter as the MRI model, clump-infall model, and collision model, respectively.
All three burst mechanisms were simulated using the same numerical hydrodynamics code FEOSAD\footnote{We note that the latest version of FEOSAD also includes dust dynamics and growth \citep{2018VorobyovAkimkin} but this feature is not employed in the present study.}. A detailed description of the code is presented in  \citet{2018VorobyovElbakyan}, with modifications relevant for modeling the MRI bursts and close encounters in 
\citet{2020Kadam} and \citet{2020VorobyovTails}, respectively. 
Here, we review only the key aspects of burst modeling.

The equations of mass, momentum, and energy transport in the thin-disk limit read
\begin{equation}
\label{cont}
\frac{{\partial \Sigma }}{{\partial t}}   + \nabla_p  \cdot 
\left( \Sigma \bl{v}_p \right) =0,  
\end{equation}
\begin{eqnarray}
\label{mom}
\frac{\partial}{\partial t} \left( \Sigma \bl{v}_p \right) +  [\nabla \cdot \left( \Sigma \bl{v}_p \otimes \bl{v}_p \right)]_p & =&   - \nabla_p {\cal P}  + \Sigma \, \bl{g}_p + \nonumber
\\ 
&+& (\nabla \cdot \mathbf{\Pi})_p,
\end{eqnarray}
\begin{equation}
\frac{\partial e}{\partial t} +\nabla_p \cdot \left( e \bl{v}_p \right) = -{\cal P} 
(\nabla_p \cdot \bl{v}_{p}) -\Lambda +\Gamma + 
\left(\nabla \bl{v}\right)_{pp^\prime}:\Pi_{pp^\prime}, 
\label{energ}
\end{equation}
where the subscripts $p$ and $p^\prime$ refer to the planar components
$(r,\phi)$  in polar coordinates, $\Sigma$ is the gas mass
surface density,  $e$ is the internal energy per surface area,  ${\cal P}$
is the vertically integrated gas pressure calculated via the ideal  equation of state as ${\cal P}=(\gamma-1) e$ with $\gamma=7/5$, $\bl{v}_{p}=v_r
\hat{\bl r}+ v_\phi \hat{\bl \phi}$  is the gas velocity in the disk plane, and is $\nabla_p=\hat{\bl r} \partial / \partial r + \hat{\bl
\phi} r^{-1} \partial / \partial \phi $ the gradient along the planar coordinates of the disk.  

The gravitational acceleration in the disk
plane,  $\bl{g}_{p}=g_r \hat{\bl r} +g_\phi \hat{\bl \phi}$, takes into account disk self-gravity found by solving for the Poisson integral \citep[for details see][]{2010VorobyovBasu} and the
gravity of the central protostar when formed. Turbulent viscosity is
taken into account via the viscous stress tensor  $\mathbf{\Pi}$, the expression for which can be found in \citet{2010VorobyovBasu}. We parameterized the
magnitude of kinematic viscosity $\nu=\alpha c_{\rm s} H$  using the $\alpha$-prescription of \citet{1973ShakuraSunyaev}, where $c_{\rm s}$ is the sound speed  calculated using the disk midplane temperature (introduced below) and $H$ is the disk vertical scale height. In the clump and collision models $\alpha$ is a constant of time and space, while in the MRI-model we use an adaptive $\alpha$-value described in more detail later in the text. 

The cooling rate per surface area is \citep{2016DongVorobyov}
\begin{equation}
\Lambda=\frac{8\tau_{\rm P} \sigma T_{\rm mp}^4 }{1+2\tau_{\rm P} + 
{3 \over 2}\tau_{\rm R}\tau_{\rm P}},
\end{equation}
where $T_{\rm mp}={\cal P} \mu / {\cal R} \Sigma_{\rm g}$ is the midplane
temperature,  $\mu=2.33$ is the mean molecular weight,  $\cal R$ is the
universal  gas constant, $\sigma$ is the Stefan-Boltzmann constant, 
$\tau_{\rm R}$  and $\tau_{\rm P}$ are the  Rosseland and Planck
optical depths to the disk midplane. We use the Planck and Rosseland mean opacities of \citet{2003SemenovHenning}.
The heating function per surface area of the disk is expressed as
\begin{equation}
\Gamma=\frac{8\tau_{\rm P} \sigma T_{\rm irr}^4 }{1+2\tau_{\rm P} + {3 \over 2}\tau_{\rm R}\tau_{\rm
P}},
\end{equation}
where $T_{\rm irr}$ is the irradiation temperature at the disk surface 
determined from the stellar and background black-body irradiation as
\begin{equation}
T_{\rm irr}^4=T_{\rm bg}^4+\frac{F_{\rm irr}(r)}{\sigma},
\label{fluxCS}
\end{equation}
where $F_{\rm
irr}(r)$ is the radiation flux (energy per unit time per unit surface
area)  absorbed by the disk surface at radial distance  $r$ from the
central star. The latter quantity is calculated as 
\begin{equation}
F_{\rm irr}(r)= \frac{L_\ast}{4\pi r^2} \cos{\gamma_{\rm irr}},
\label{fluxF}
\end{equation}
where $\gamma_{\rm{irr}}$ is the incidence angle of radiation arriving at the disk surface (with respect to the normal) at radial distance $r$ \citep[see][for details]{2010VorobyovBasu}. The stellar luminosity $L_{\ast}$ is the sum of the accretion and photospheric luminosities. The former is computed as  
\begin{equation}
  L_{\rm {\ast,accr}}=  {1\over 2} {G M_{\ast} \dot{M}  \over R_{\ast} },
  \label{Laccr}
\end{equation}
where $M_\ast$ and $\dot{M}$ are the stellar mass and mass accretion rate, respectively, and $G$ is the gravitational constant. The stellar radius $R_\ast$ and
photospheric luminosity $L_{\rm{\ast,ph}}$ due to gravitational
compression and deuterium burning in the stellar interior are
calculated using the stellar evolution tracks obtained with the STELLAR code of \citet{2008YorkeBodenheimer}. We note that Equation~(\ref{Laccr}) does not take into account the possible variations of the accretion luminosity caused by the processes at the stellar surface (e.g., a fraction of accretion energy absorbed by the star, \citet{2017BaraffeElbakyan}), which may affect the burst light curves but are of less significance for the analysis of disk kinematic signatures.


FEOSAD starts simulations from a collapsing pre-stellar core in the form of a flattened pseudo-disk, which is expected in the presence of rotation and large-scale magnetic fields \citep{1997Basu}.  The inner regions spin  up  and  a  circumstellar  disk  forms  when  the  inner  infalling  layers  hit  the  centrifugal  barrier near  the  inner computational boundary.   The  material  passing  through  the inner boundary  forms  the  growing  central  star.   The infalling core continues to land at the outer edge of the circumstellar disk (a reasonable approximation according to \citet{2009Visser})  until the core depletes.  The infall rates on the circumstellar disk are in agreement with analytic collapse models \citep{2010Vorobyov}.

The main model characteristics are listed in Table~\ref{tab:1}. The parameters of pre-stellar cloud cores with distinct masses $M_{\rm core}$, initial temperatures $T_{\rm init}$, and ratios of rotational-to-gravitational energy $\beta$  were chosen to produce disks with different characteristics depending on the particular burst mechanism. For instance, to study the bursts triggered by clump infall, we set a more massive pre-stellar core with a higher rate of rotation to produce a massive and extended disk prone to fragment. The initial temperature, which is also the temperature of external stellar irradiation in our models, is also lower to promote disk fragmentation. The other two models are characterized by lower $M_{\rm core}$ and $\beta$ but higher $T_{\rm init}$ to study the bursts triggered by the MRI and stellar encounters,  In these later models the interference from the clump-infall mechanism is minimized.  The simulations continued for up to 500~kyr to capture the entire embedded and early T~Tauri stages of disk evolution, but we focus here only on short time periods with representative bursts.

The MRI bursts considered here were modelled in detail in \citet{2020Kadam} with the help of the adaptive $\alpha$-parameterization of turbulent viscosity following the method laid out in \citet{2014BaeHartmann}. In the MRI model, known also as the layered disk model \citep{2001Armitage}, the $\alpha$-value is weighed according to the thickness of the MRI-dead and MRI-active vertical columns of the disk. The MRI-active column and the corresponding $\alpha$-value are set equal to $\Sigma_{\rm a}=100$~g~cm$^{-2}$  and $\alpha_{\rm max}=0.01$, respectively.  The $\alpha$-value of the MRI-dead column $\Sigma_{\rm d}$ is set to $\alpha_{\rm d}=10^{-5}$. 
The adaptive $\alpha$-value then reads
\begin{equation}
\alpha={\Sigma_{\rm a} \alpha_{\rm max} + \Sigma_{\rm d} \alpha_{\rm d} \over \Sigma},
\label{alphaV}
\end{equation}
where the total surface density of disk is $\Sigma= \Sigma_{\rm a} + \Sigma_{\rm d}$. Equation~(\ref{alphaV}) indicates that the outer disk regions with low surface densities are MRI-active, while in the innermost disk regions (where the vertical column of gas greatly exceeds 100~g~cm$^{-2}$) the $\alpha$-value effectively reduces to $\alpha \simeq10^{-5}$. These latter regions of reduced viscous mass transport constitute a 'dead' zone where matter accumulates while being transported inwards from the disk outer regions by the combined action of gravitational and/or viscous torques. 

The MRI burst is triggered when the gas temperature in the innermost disk regions exceeds a threshold value of $T_{\rm crit}=1300$~K. Above this value, thermal ionization of alkaline metals sets in and the dead zone becomes active over a short period of time. This transition is implemented by a sudden increase in the $\alpha$-value throughout the entire disk vertical column to a peak value of $\alpha_{\rm max}=0.1$. The active state continues for as long as the disk temperature stays above 1300~K.  We note that the peak value of $\alpha_{\rm max}=0.1$ is higher in the disk regions directly involved in the burst than in otherwise MRI-active regions with $\alpha_{\rm max}=0.01$ (e.g., outer disk with low column density). This choice is motivated by numerical magnetohydrodynamics simulations of \citet{2020Zhu} suggesting that the $\alpha$-value in the innermost disk regions during the MRI burst can exceed notably the typical value of 0.01 for MRI-active disks in the non-burst state \citep{2018Yang}.

\begin{table*}
\center
\caption{\label{tab:1}Model parameters}
\begin{tabular}{cccccccccc}
\hline 
\hline 
Burst & Burst & $M_{\mathrm{core}}$ &  $\beta$ & $T_{\rm init}$ & $M_\ast$ & $M_{\rm disk}$ & $T_{\rm crit}$ & $\alpha_{\rm{max}}$  & $r_{\rm per}$    \tabularnewline
number & type & [$M_{\odot}$]  &  & [K] & [$M_{\odot}$] & [$M_{\odot}$] & [K] & & [au] \tabularnewline
\hline 
1 & MRI & 1.0  &  $1.17\times10^{-3}$ & 15 &  0.63 & 0.357 & 1300 & 0.01--0.1 & -  \tabularnewline
2 & Clump infall & 1.1  & $6.12\times10^{-3}$ & 10 & 0.79 & 0.195 & - & 0.01 & - \tabularnewline
3 & Clump infall & 1.1  & $6.12\times10^{-3}$ & 10 & 0.79 & 0.179 & - & 0.01 & - \tabularnewline
4 & Collision & 0.66  & $2.2\times10^{-3}$ & 15 & 0.47 & 0.056 & - & 0.01 & 82.4 \tabularnewline
5 & Collision & 0.66  & $2.2\times10^{-3}$ & 15 & 0.47 & 0.056 & - & 0.01 & 75.3 \tabularnewline
\hline 
\end{tabular}
\center{ \textbf{Notes.} 
 $M_{\mathrm{core}}$ is the initial core mass,  $\beta$ is the ratio of rotational to gravitational energy, $T_{\rm init}$ is the initial temperature of the core, $M_\ast$ and $M_{\rm disk}$ are the stellar and disk masses at the time instance of the burst,  $T_{\rm crit}$ is the threshold temperature for MRI ignition, $\alpha_{\rm max}$ is { the fixed $\alpha$-parameter value for all the models except the MRI model where the values refer to the range of maximum $\alpha$'s utilized throughout the disk}, and $r_{\rm per}$ is the periastron distance of the intruder.}
\end{table*}


Accretion bursts caused by infall of gaseous clumps were studied by \citet{2015VorobyovBasu}. This mechanism requires massive gravitationally unstable disks prone to fragment. Gravitational interaction between the clumps and also between the clumps and spiral arms causes the clumps to preferentially migrate inward. Fast inward migration followed by tidal destruction delivers large amounts of matter to the inner disk regions causing an accretion burst. This mechanism has recently been confirmed to operate also in disks around massive stars \citep{2017MeyerVorobyov}. In this paper, we used the high-resolution models of \citet{2018VorobyovElbakyan} which were able to resolve the internal structure of inward-migrating clumps as they tidally lose their envelope and produce the burst. 

\begin{figure}
\begin{centering}
\includegraphics[width=1\columnwidth]{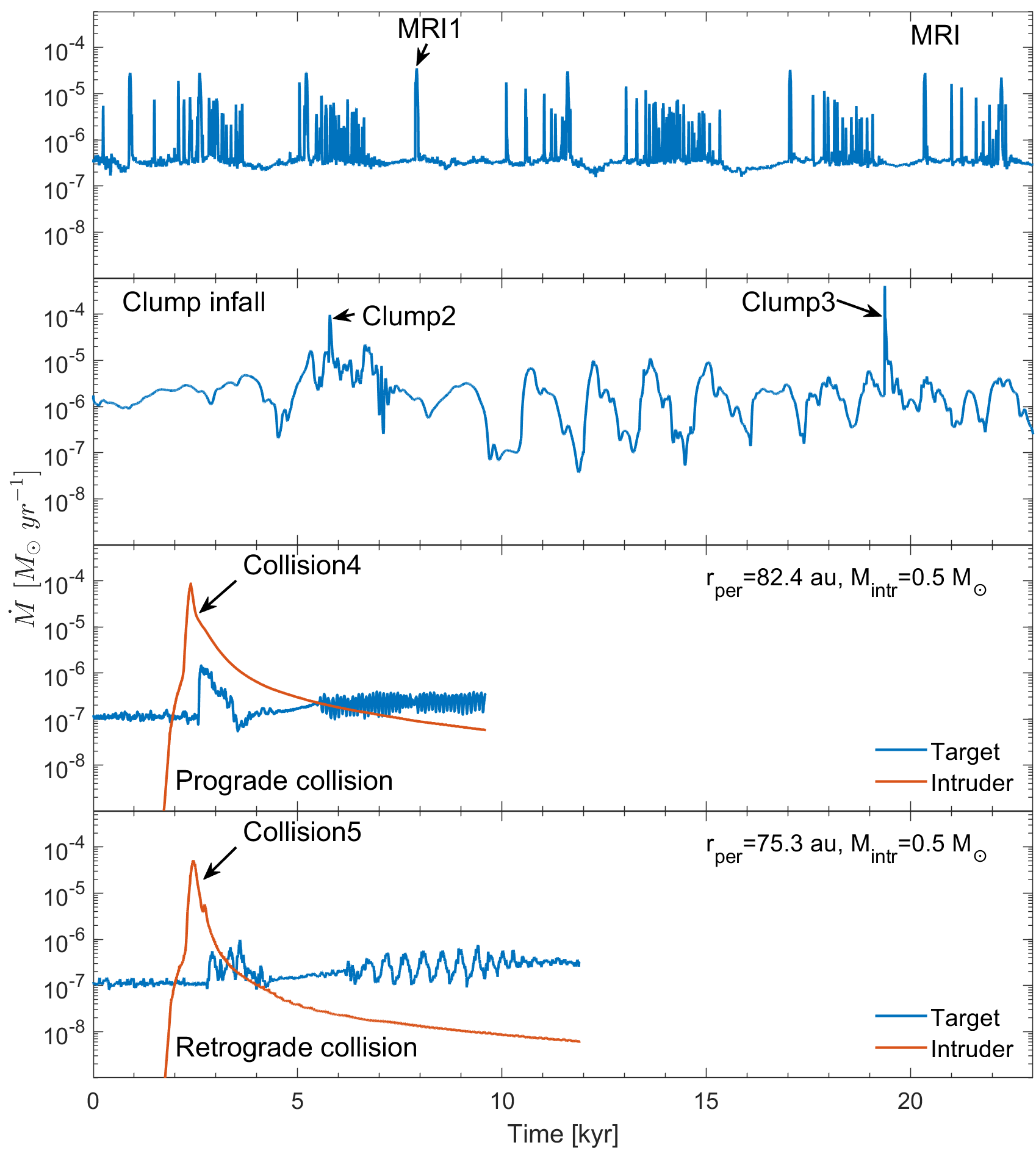}
\par\end{centering}
\caption{\label{fig:1} Accretion rate vs. time in the MRI model (top panel), clump-infall model (second panel), and collision model (third and forth panels). The collision model includes the cases of prograde (third panel) and retrograde (forth panel) collisions.  In particular, the red and blue lines present the mass accretion rates on the intruder and target, respectively. The time is reset to zero for each chosen interval of disk evolution. The arrows indicate the bursts that are investigated in more detail later in the text.}
\end{figure}

To study the bursts caused by close encounters, we use the model recently presented in \citet{2020VorobyovTails}. In this collision model, a diskless star is set on an encounter trajectory with a protoplanetary disk around a sub-solar mass star. The intruder star is allowed to accrete matter from its surroundings and its accretion luminosity is taken into account when computing the thermal balance of the system. 
We modified the accretion rate calculations of the intruder star using the following algorithm. First, we calculate the Hill radius of the intruder ($R_{\rm H}$) and consider only the grid cells within $R_{\rm H}$ that satisfy the following criterion
\begin{equation}
    E_{\rm kin} + E_{\rm gr} < 0,
    \label{enIntr}
\end{equation}
where $E_{\rm kin}$ is the kinetic energy of motion in the frame of reference of the intruder and $E_{\rm gr}$ is the gravitational energy in its gravitational field. The above equation states that the material has to be on a bound orbit to be accreted by the intruder. We then calculate the mass of gas that is accreted by the intruder during one time step $dt$ as
\begin{equation}
    \Delta M = \sum_i {\cal D}_i {\cal F}_i \, \Sigma_i \, dS_i,
\end{equation}
where the summation is performed over all cells within the Hill radius subject to condition~(\ref{enIntr}), $dS_i$ is the surface area of a given cell, $\Sigma_i$ is the surface density in this cell, and ${\cal F}_i$ is the fraction of accreted material in this cell defined as
\begin{equation}
    {\cal F}_i = dt \, {\Omega_{\rm K,i} \over 2 \pi}.
    \label{fracAcr}
\end{equation}
Here, $\Omega_{\rm K,i}$ is the Keplerian velocity of a given cell in the frame of reference of the intruder. Equation~(\ref{fracAcr}) implies that the intruder accretes all material in a given cell on a Keplerian time scale, but this never happens in reality because Equation~(\ref{enIntr}) may be violated and new material is captured by the intruder as it passes by. To take the finite disk thickness into account, we multiply the accreted mass $\Delta M$  by the following factor
\begin{equation}
{\cal D}_i  = \left\{ \begin{array}{ll} 
   {H_i^{-1}\sqrt{R_{\rm H}^2 - r_{i}^2}},  &\,\,\,  \mbox{if} \,\,\,  R_{\rm H}^2 - r_{i}^2 < H_i^2, \\ 
   1.0,  & \,\,\, \mbox{otherwise}, 
   \end{array} 
   \right. 
   \label{function} 
\end{equation}
where $H_i$ is the disk vertical scale height in a given cell and $r_{\rm i}$ is the radial distance from the intruder to a given cell. We note that the value of ${\cal D}_i$ is expected to be unity for the considered masses and periastron distances of the intruder, but may be smaller than unity in a general case.  Once the accreted mass $\Delta M$ is calculated, the mass of the intruder, its velocity, the surface densities and velocities of affected grid cells within the Hill radius are updated to conserve the mass and momentum. The ideas described above were taken from \citet{2010ForganRice} and \citet{2012KleyNelson}. We considered also other accretion prescriptions \citep[e.g.,][]{2010FederrathBanerjee}, but found them less realistic in our context because they led to premature dissipation of the disk captured by the intruder. Finally, we note that Equations~(\ref{mom}) and (\ref{energ}) were modified take the gravitational potential and luminosity of the intruder into account. The computations were performed in the non-inertial frame of reference of the target star by introducing the so-called indirect potential \citep[see for details][]{2017VorobyovSteinrueck}. The $\alpha$-parameter in the collision and clump infall models is set equal to a constant value of  $10^{-2}$ to exclude  MRI-triggered bursts.

In this paper, we consider accretion bursts caused by both prograde and retrograde encounters in the plane of the target disk.  The disk was evolved to an age of 0.5~Myr to guarantee that the other considered bursts mechanisms are unlikely to operate. Indeed, the disk is axisymmetric and gravitationally stable at this late stage, which excludes the clump-infall mechanism. We also make sure that the disk density and temperature at this time instance are insufficient to produce the MRI-triggered burst without external interference. Typically, accretion of matter on the intruder, as it passes through the disk, causes a burst. We note, however, that perturbations produced by the intruder can trigger disk fragmentation in the disk of the target \citep[e.g.][]{2010Thies}. Besides, the mass inflow to the inner disk caused by the close passage of the intruder can also trigger the MRI burst in the disk of the target. The situation can therefore become quite complicated in the case of encounter-triggered bursts. Here, we consider in detail only the primary burst of the intruder star and leave the investigation of secondary bursts for a follow-up study.

All bursts mechanisms were considered on the polar grid ($r,\phi$) in the thin-disk geometry. The radial grid is logarithmically spaced, while the azimuthal grid is equally spaced. To model the MRI bursts, the inner disk boundary is set at $r=0.4$~au to capture the innermost disk regions where the MRI is supposed to operate. The number of grid cells is $512\times512$, which corresponds to a numerical resolution of 0.02~au at one astronomical unit. In the clump-infall model, the inner boundary is set at 15~au and the number of grid cells is $1024\times1024$. This choice allowed us to attain a sub-au numerical resolution up to a radial distance of 150~au, i.e., in the disk regions of interest where clumps form and migrate. In the close-encounter model, the inner boundary is set at 2~au and the number of grid cells is $512\times512$. 

\begin{figure}
\begin{centering}
\includegraphics[width=1\columnwidth]{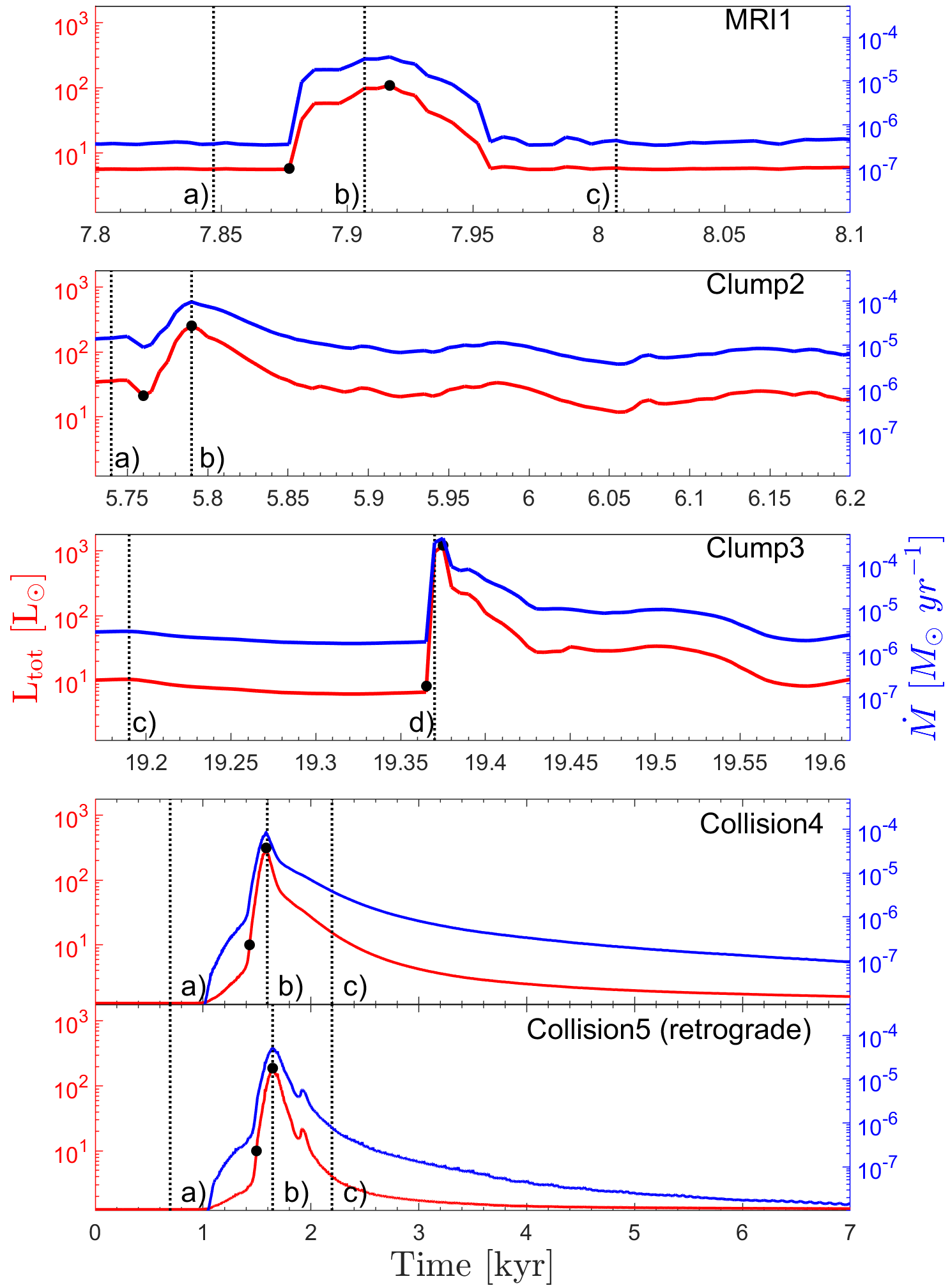}
\par\end{centering}
\caption{\label{fig:2} Temporal evolution of the accretion rate (blue lines) and total luminosity (red lines) for the five bursts indicated with arrows in Figure~\ref{fig:1}.  The vertical black dotted lines marked with letters indicate the time instances studied in more detail further in this section.}  
\end{figure}

The inner unresolved region of the disk is replaced with a sink cell and matter is allowed to flow both directions -- from the disk to the sink and vice versa \citep[for details see][]{2019Kadam}. 
The mass accretion rate in the MRI and clump-infall models is calculated as the mass passing through the sink cell per unit time. In the collision model, we focus on the mass accretion on the intruder and not on the target star (the latter is much smaller than the former). In this model, the accretion rate on the intruder is calculated following the method described by Equations~(\ref{enIntr})-(\ref{function}).

We note that the calculated mass accretion rate best represents the protostellar accretion rate in the MRI model. In the collision model, the mass accretion rate on the intruder star is subject to model assumptions laid out by Equations~(\ref{enIntr})-(\ref{function}) and may change if other accretion models were used. In the clump infall model, physical mechanisms operating inside 15 au may also modify the calculated mass accretion rate.  Nevertheless, these model limitations are not expected to affect notably the global kinematics of protoplanetary disks undergoing accretion bursts. In this case, using one code to model different burst mechanisms presents a clear advantage, allowing us to  eliminate the uncertainty that may be introduced by adopting different disk physics and numerical methods.

\begin{figure}
\begin{centering}
\includegraphics[width=1\columnwidth]{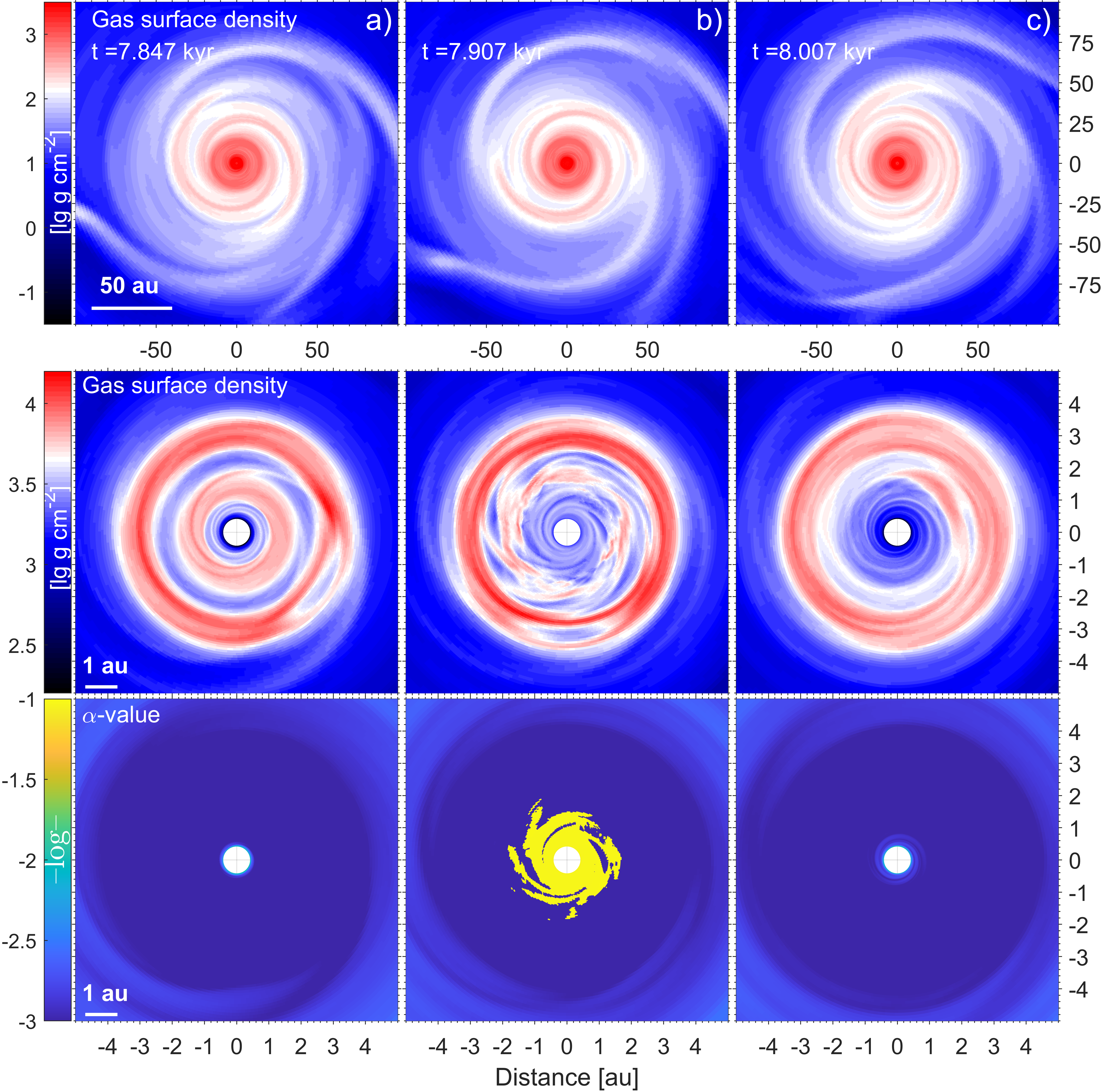}
\par\end{centering}
\caption{\label{fig:3} Disk properties  before (left column), during (middle column), and after (right column) the MRI burst (MRI1). In particular, the columns from left to right correspond to the time instances marked in the top panel of Figure~\ref{fig:2} with the vertical dotted lines a), b), and c), respectively. \textbf{Top row:} Gas surface density maps in the inner 200$\times$200~au$^2$ box of the disk.
\textbf{Middle row:} Gas surface density in the inner 10$\times$10~au$^2$ part of the disk. \textbf{Bottom row:} The value of $\alpha$-parameter in the inner 10$\times$10~au$^2$ part of the disk. The white circles in the coordinate center represent the sink cell.}
\end{figure}


\section{Kinematic signatures of accretion bursts}
\label{Kinematics}
In this section, we provide a detailed study of protoplanetary disks subject to accretion and luminosity bursts. The focus is put on the analysis of disk kinematics. In particular, we will search for any signatures in the disk rotation curves and velocity channel maps that may help us to distinguish between different burst-triggering mechanisms.

\subsection{Considered accretion bursts} 
\label{IndBursts}

We start by illustrating the accretion and luminosity bursts considered in our work.
Figure~\ref{fig:1} presents the mass accretion rate versus time for several representative time intervals of disk evolution. We do not show the entire computed disk evolution (which is much longer in the MRI and clump-infall models) because we focus in this study on  the kinematic signatures of individual bursts rather than on the collective burst properties.   Panels from top to bottom show the MRI model (first panel), clump-infall model (second panel), and collision model (third and forth panels). 
The collision model is represented by two panels for the accretion rate on the intruder in the prograde and retrograde encounters. The target star exhibits bursts of a much smaller amplitude than is typical of most FU Orionis-type eruptions \citep[see also][]{2010ForganRice}. 
The MRI and clump-infall models show  highly variable accretion with multiple bursts, while the mass accretion rate on the intruder in the collision model is characterized by one burst that occurs during the closest approach between the intruder and target stars ($t\approx2.5$~kyr).  High variability in the clump-infall model is caused by the perturbing influence of the clump that orbits the star at about 25~au. 

We have chosen several representative accretion bursts indicated by arrows in Figure~\ref{fig:1}, the total luminosities and mass accretion rates of which are displayed on shorter time intervals in Figure~\ref{fig:2}. Clearly, the considered mechanisms produce a variety of light curves with different peak luminosities, burst durations, rise and decay times.
We calculated the burst magnitudes as
\begin{equation}
    \Delta m = - 2.5 \log_{10} {L_{\rm peak} \over L_{\rm base}},
\end{equation}
where $L_{\rm peak}$ and $L_{\rm base}$ are the peak and preburst total luminosities marked in Figure~\ref{fig:2} with the black circles.  Choosing the preburst luminosity for the collision-triggered bursts presents a certain difficulty because the intruder was
diskless before the collision and its total luminosity was determined exclusively by its photospheric contribution (set equal to $1~L_\odot$). As the intruder penetrates the target disk, its accretion (and total) luminosity is gradually rising. We set the preburst luminosity equal to $10~L_\odot$, which is comparable to the corresponding value in the clump-infall models. The resulting values of $\Delta m$ lie in the 2.5--3.7 limits,  except  for  the clump-infall  model  where $\Delta m$  can  reach  5.4 for the Clump3 burst. The peak luminosities of model bursts are { in the range of the known FU Orionis type bursts} (see Table~\ref{FUors} in the Appendix).
We postpone a detailed analysis of these burst characteristics for a follow-up paper and focus on the disk kinematic signatures that are associated with these bursts. The vertical dotted lines mark the time instances before, during, and after the bursts, which are chosen to investigate in detail the disk kinematics below.


\begin{figure}
\begin{centering}
\includegraphics[width=1\columnwidth]{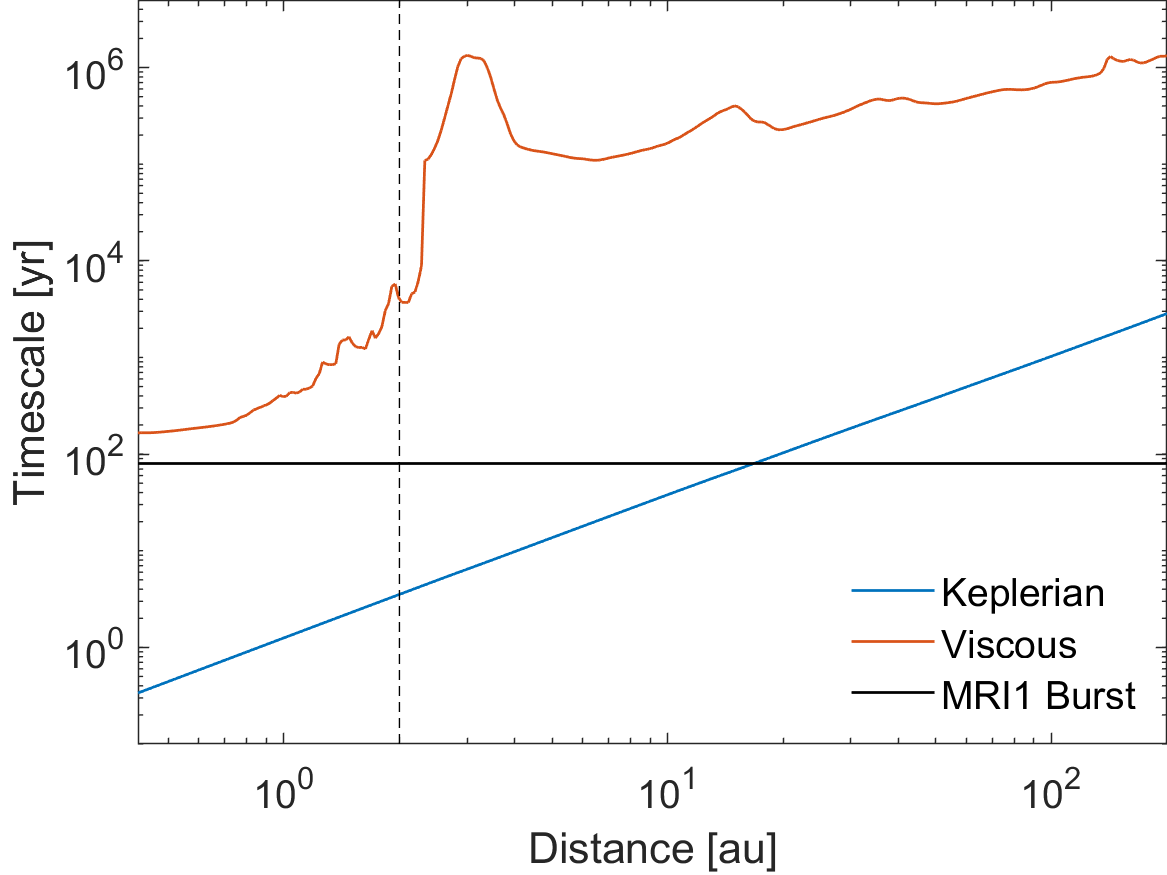}
\par\end{centering}
\caption{\label{fig:4b} Comparison of the relevant time scales in the disk during the MRI1 burst. The red and blue lines present the viscous and dynamical time scales, respectively, while the black line shows the duration of the burst. The vertical dotted line outlines the extent of the MRI-active inner disk region. }
\end{figure}

\begin{figure}
\begin{centering}
\includegraphics[width=1\columnwidth]{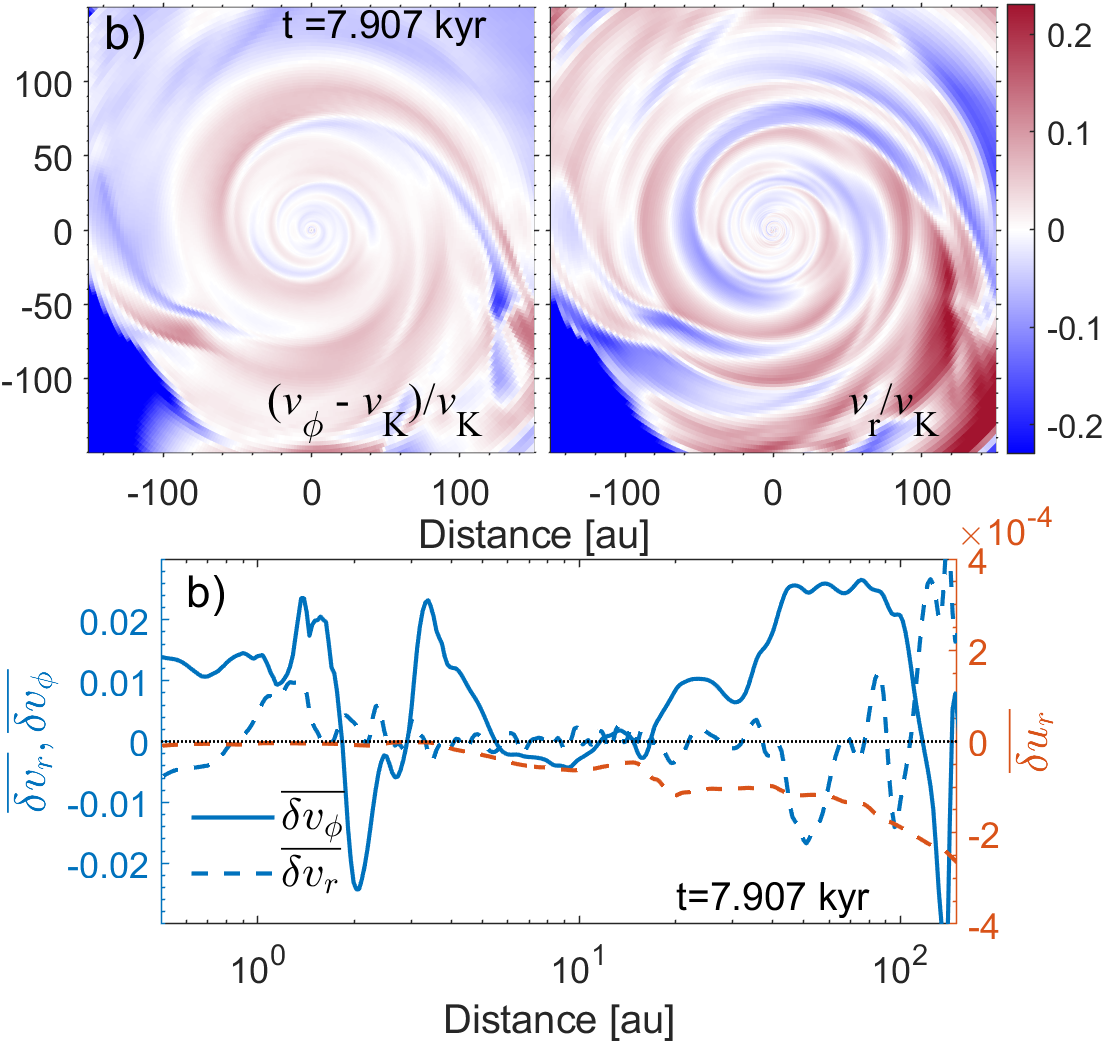}
\par\end{centering}
\caption{\label{fig:4} {\bf Top panels:} Spatial distributions of the residual azimuthal velocity $\delta v_\phi=(v_{\phi} - v_{\rm K})/v_{\rm K}$ and the ratio of radial to Keplerian velocity  $\delta v_r=v_{\rm r}/v_{\rm K}$ in the MRI model. 
The data correspond to the time instance near the peak of the MRI1 burst marked in the top panel of Figure~\ref{fig:2} with the vertical dotted line b). The white circles in the coordinate center represent the sink cell. The disk rotates counterclockwise. 
{\bf Bottom panel:} The corresponding azimuthally averaged radial profiles of residual velocity $\delta v_\phi$ and ratio $\delta v_r$.  The red dashed line shows the ratio of radial to Keplerian velocity  $\delta u_r=u_{\rm r}/v_{\rm K}$ for an idealized steady-state disk at the time instance immediately preceding the burst.}
\end{figure}

\begin{figure*}
\begin{centering}
\includegraphics[width=2\columnwidth]{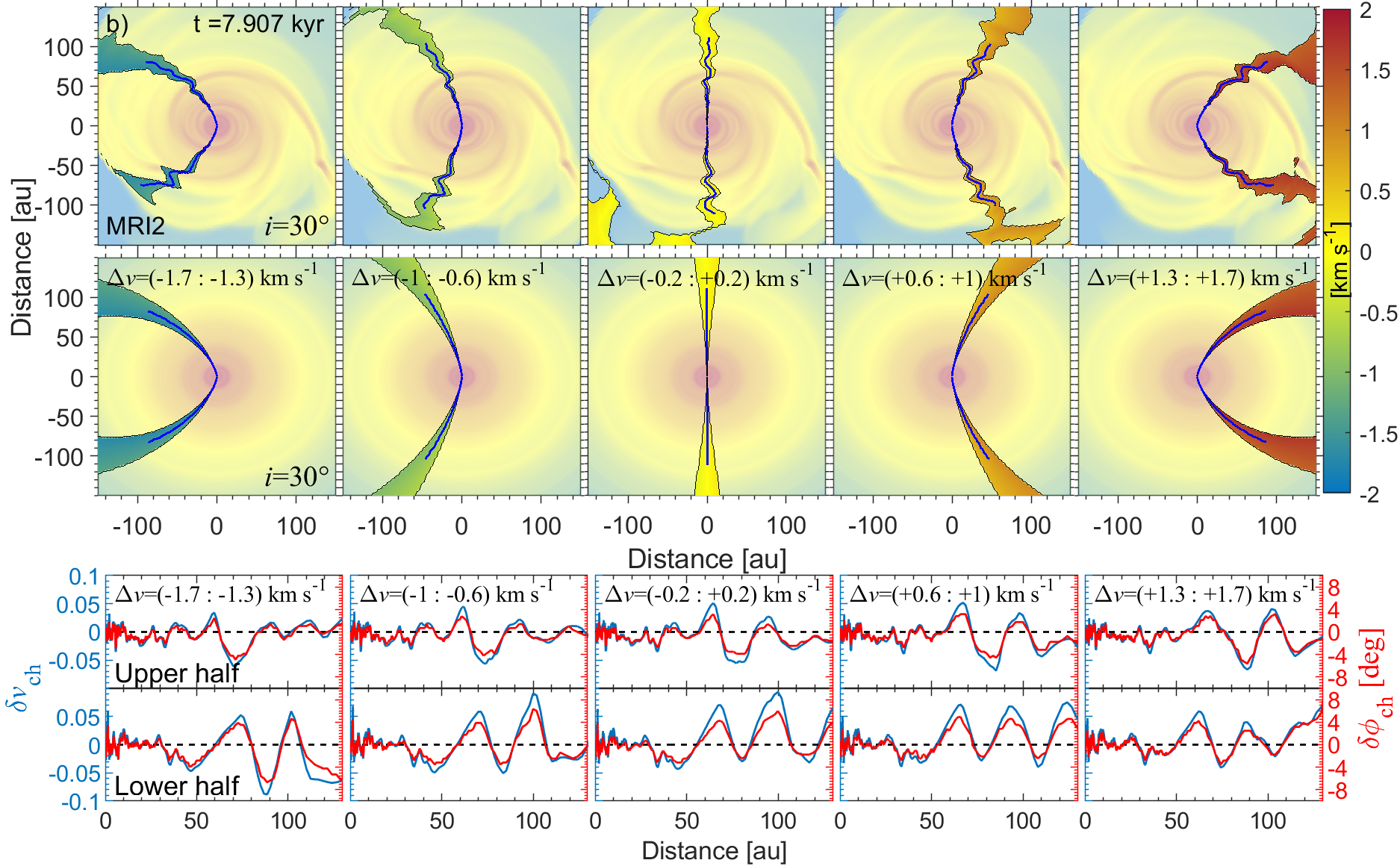}
\par\end{centering}
\caption{\label{fig:16a} Line-of-sight velocity channel maps in the disk of the MRI model near the burst maximum (top row) vs. those  of an idealized Keplerian disk (middle row).  The disks are tilted by 30$^\circ$ with respect to the horizontal axis with the upper part being further out from the observer. The color bar presents the deviation from the zero-velocity in km~s$^{-1}$. Each channel shows a velocity interval of 0.4~km~s$^{-1}$.  The corresponding gas surface density maps are plotted in pale palette for convenience. The rotation is counterclockwise.  The bottom row displays the deviations in the position angle ($\delta \phi_{\rm ch}$) and velocity ($\delta v_{\rm ch}$) of the center of each velocity channel in the MRI model with respect to the center of the corresponding channel in the Keplerian disk. The centers of the model and  Keplerian channels are shown by the blue curves in the top and middle panels for convenience.  The values for the upper and lower halves of the disk are shown separately.  
}
\end{figure*}

\subsection{MRI-triggered bursts}

Figure~\ref{fig:3} presents the two-dimensional disk properties before, during, and after the MRI burst for the light curve shown in the top panel of Figure~\ref{fig:2}. The corresponding time instances are indicated by the vertical dotted lines and marked with letters a), b), and c) in both Figures~\ref{fig:2} and \ref{fig:3}.  The top and middle panels show the gas surface density with different spatial resolution, while the bottom panel presents the adaptive $\alpha$-value. The global disk structure is dominated by spiral arms indicating that the disk is gravitationally unstable. Gravitational instability is a mechanism that, alongside with turbulent viscosity, helps to bring matter in the disk innermost regions and trigger the MRI \citep{2009ZhuHartmannGammie}. The inner disk structure is dominated by gaseous rings that form in the dead zone \citep{2019Kadam}. During the burst (middle column) the inner ring becomes MRI-unstable and falls on the star, thus producing a burst. The development of the MRI is evident in the bottom panel showing that the $\alpha$-parameter rises to a peak value of 0.1 during the burst. About 2.9~$M_{\rm J}$ of gas mass is accreted during the burst from the inner 1.8~au. After the burst, the inner region with a radius of several astronomical units features a gap, which slowly fills in with matter as the disk evolves and recovers from the burst.

Figure~\ref{fig:4b} compares the relevant  time scales in the disk during the MRI burst. In particular, the dynamical (Keplerian) timescale $\tau_{\rm K}= 2\pi r / v_{\rm K}$ is shown by the blue line, while the black line presents the burst duration $\tau_{\rm bst}$ calculated by defining the baseline that connects the two time points of equal luminosity on both sides of the luminosity peak. The red line shows the viscous timescale defined as $\tau_{\rm visc}=r^2/\nu$, where we azimuthally averaged the kinematic viscosity $\nu$ to derive the radial profile. {We note that the viscous timescale is therefore 
inversely dependent on the $\alpha$-parameter and the sound speed in the disk}. In the disk region engulfed by the burst (to the left from the vertical dotted line), $\tau_{\rm visc}\ga \tau_{\rm bst}$ and both timescales are much longer than the dynamical timescale. The fact that $\tau_{\rm visc}\gg \tau_{\rm bst}$ in the bulk of the disk and $\tau_{\rm visc}$ approaches $\tau_{\rm bst}$ in the innermost disk regions affected by the burst supports its viscous origin. At the same time $\tau_{\rm bst}$ remains shorter than $\tau_{\rm visc}$, which reflects a self-regulating nature of the MRI burst -- the burst terminates when the inner disk becomes depleted of matter owing to elevated viscous mass transport.

To analyze the disk kinematics during the burst, we calculated the residual {azimuthal} velocity  and the ratio of radial to Keplerian velocity  as
\begin{eqnarray}
\delta v_\phi &=& (v_\phi-v_{\rm K})/v_{\rm K}, \\
\delta v_r &=& v_r/v_{\rm K}.
\end{eqnarray}
When calculating the Keplerian velocity $v_{\rm K}$ we also took into account the disk mass that lies interior to a given radial distance. The two-dimensional distribution of these quantities is shown in  the upper panel of Figure~\ref{fig:4} at a spatial scale that captures the entire disk. The time instance near the peak of the burst is chosen.  The residual velocity $\delta v_\phi$ and the ratio $\delta v_r$ exhibit a spiral-like spatial pattern that is caused by global perturbations from the spiral density waves traversing the disk. The amplitude of these perturbations is smaller in the inner disk regions where the waves are weakest (due to increasing temperature and shear).
The bottom panel in Figure~\ref{fig:4} presents the azimuthally averaged residual { azimuthal} velocity $\overline{\delta v_\phi}$ and the ratio of radial to Keplerian velocity $\overline{\delta v_r}$. When calculating the average, we mass-weighted the radial and azimuthal velocities to diminish the input from the disk regions where little matter is localized. The azimuthally averaged patterns are highly irregular, but in general are of rather small amplitude, not exceeding a few per cent. 
In particular, $\overline{\delta v_r}$ is notably smaller than $\overline{\delta v_\phi}$ and the former is negative in the inner 1~au, reflecting the infall of matter caused by the MRI.

Furthermore, we calculated the ratio of radial to Keplerian velocity ${\delta u_r}=u_{\rm r}/v_{\rm K}$ for a steady-state disk with the mass transport rate defined as   
\begin{equation}
   \dot{M}_{\rm st} =  3 \pi \mu \left[ 1 - \left({R_\ast \over r}\right)^{0.5}  \right]^{-1},
    \label{steadystate}
\end{equation}
where $\mu$ is the dynamic viscosity and $R_\ast=3~R_\odot$ is the radius of the central star. The radial velocity $u_{\rm r}$ of such an idealized disk can be derived noting that the mass transport rate is $\dot{M}_{\rm st}=2 \pi r u_{\rm r} \Sigma$. For $\mu$  in Equation~(\ref{steadystate}) we took the azimuthally averaged quantities from the MRI model at the time immediately preceding the burst, $t=7.847$~kyr (the highly perturbed disk during the MRI burst cannot be described by a steady-state approach). The resulting deviation ${\delta u_r}=u_r/v_{\rm K}$ is shown with the red dashed line in the bottom panel of Figure~\ref{fig:4}.
The values of ${\delta u_r}$ for the steady-state disk are found to be much smaller than the variations in $\overline{\delta v_r}$ for the real disk during the MRI burst.
In fact, ${\delta u_r}$ (the red dashed line) do not exceed $2\times10^{-4}$ by absolute value in the inner 200~au, while $\overline{\delta v_r}$ (the blue dashed line) can be as large as $\pm0.01$. Our calculated values of $\delta u_r$ are in agreement with analytical estimates for a steady-state disk \citep{1998Hartmann}
\begin{equation}
    \delta u_r \simeq {3 \over 2} \alpha \left( {H \over r} \right)^2,
\end{equation}
for a typical disk aspect ratio of $H/r=0.1$.
{ It is worth noting that mass transport in real disks can be more complex than predicted by Equation~(\ref{steadystate}) and disk's outer parts can instead spread out. Thus, $\delta u_r$ is an order of magnitude estimate, which simply demonstrates that our model velocity variations are much larger than this estimate.}

The top row in Figure~\ref{fig:16a} compares the velocity channel maps for Burst~2 near its maximum  with the velocity channel maps of an axisymmetric unperturbed Keplerian disk (middle row).  When calculating the velocity of the Keplerian disk we also take the enclosed disk mass into account. The gas surface density of the Keplerian disk is obtained by azimuthally averaging the surface density of our model disk near the peak of the burst. For convenience we also plot the gas surface density of the disk (the color map for the surface density is chosen arbitrarily for better visual representation). 
Each channel represents a velocity interval (along the line of sight) of 0.4~km~s$^{-1}$. It is assumed that the disk is tilted by 30$^\circ$ relative to the horizontal axis with the upper part being further out from the observer. 
The model velocity channel maps show 'kinks' and 'wiggles', similar to those found by \citet{2020Hall} for a gravitationally unstable disk. These features are caused by the spiral density waves that traverse and perturb  the disk. They are absent in the idealized Keplerian disk. In general there are no strong deviations from the Keplerian velocity channel maps for the inner several tens of astronomical units of the disk because the spiral pattern is weakened there. Notable deviations occur only in the outer disk regions in the vicinity of strong spiral arms. We conclude that the velocity channel maps can help to reveal the presence of gravitational instability as a possible MRI-assisting mechanism but not the MRI burst itself. Whether or not these velocity fluctuations can be detected is uncertain  and the line radiative transfer simulations are needed to address this point.  However, the spiral arms causing these deviations are likely to be visible in the dust continuum emission \citep[see, e.g.,][]{2016DongVorobyov}.

From the observations of dense gas tracers toward a Keplerian rotating disk, the  blueshifted and redshifted velocity channels at the same velocity offset (from the systemic velocity) should present reflection symmetric with respect to the stellar position. This effect is illustrated in the middle row of Figure~\ref{fig:16a}. Quantifying the deviation from such reflection symmetric may provide the degree of deviation from the Keplerian rotation. To perform such an analysis, we calculate the deviations in the position angle and velocity of the center of each velocity channel in the MRI model with respect to the center of the corresponding channel in the Keplerian disk. The corresponding centers of the velocity channels are shown in the top and middle panels of Figure~\ref{fig:16a} by the solid blue curves. The corresponding deviations are defined as
\begin{eqnarray}
\delta \phi_{\rm ch} (r) &=& \phi_{\rm model} - \phi_{\rm K},  
\label{angle_ch}
 \\
\delta v_{\rm ch} (r) &=& {v^{\rm l.o.s}_{\rm model}  - v^{\rm l.o.s}_{\rm K}  \over v^{\rm l.o.s}_{\rm K}},
\label{vel_ch}
\end{eqnarray}
where $\phi_{\rm model}$ and $\phi_{\rm K}$ are the position angles of the center of the model and Keplerian channel maps, which are calculated by taking a cut along the circumference with a fixed radial distance $r$. We note that there are two position angles corresponding to the upper and lower halves of the projected disk and the position angles are counted counterclockwise from the positive segment of the $x$-axis. The quantities $v^{\rm l.o.s}_{\rm model}$ and $v^{\rm l.o.s}_{\rm K}$ are the line-of-sight velocities that are calculated at the centers of the model and Keplerian velocity channels (blue curves in the top and middle panels, respectively). If the velocity field in the MRI model were unperturbed by the spiral pattern and pressure gradients, then a Keplerian disk would effectively be retrieved and both quantities $\delta \phi_{\rm ch}$ and $\delta v_{\rm ch}$  would be negligibly small.

The resulting values of $\delta \phi_{\rm ch}$ and $\delta v_{\rm ch}$ as a function of radial distance are plotted in the bottom panel of Figure~\ref{fig:16a}. The deviations in the position angle $\delta \phi_{\rm ch}$ do not exceed ten degrees and stay mostly within a few degrees. The corresponding relative deviations in the line-of-sight velocity $\delta v_{\rm ch}$ also stay within several percent. The biggest deviations are found in the outer disk regions where the spiral pattern is the strongest. However, nonzero deviations can also be noted in the innermost disk regions, which is not evident in the top and middle panels of Figure~\ref{fig:16a} because of the narrowing channel maps near the coordinate center.

\subsection{Bursts triggered by clump-infall}
\begin{figure}
\begin{centering}
\includegraphics[width=1\columnwidth]{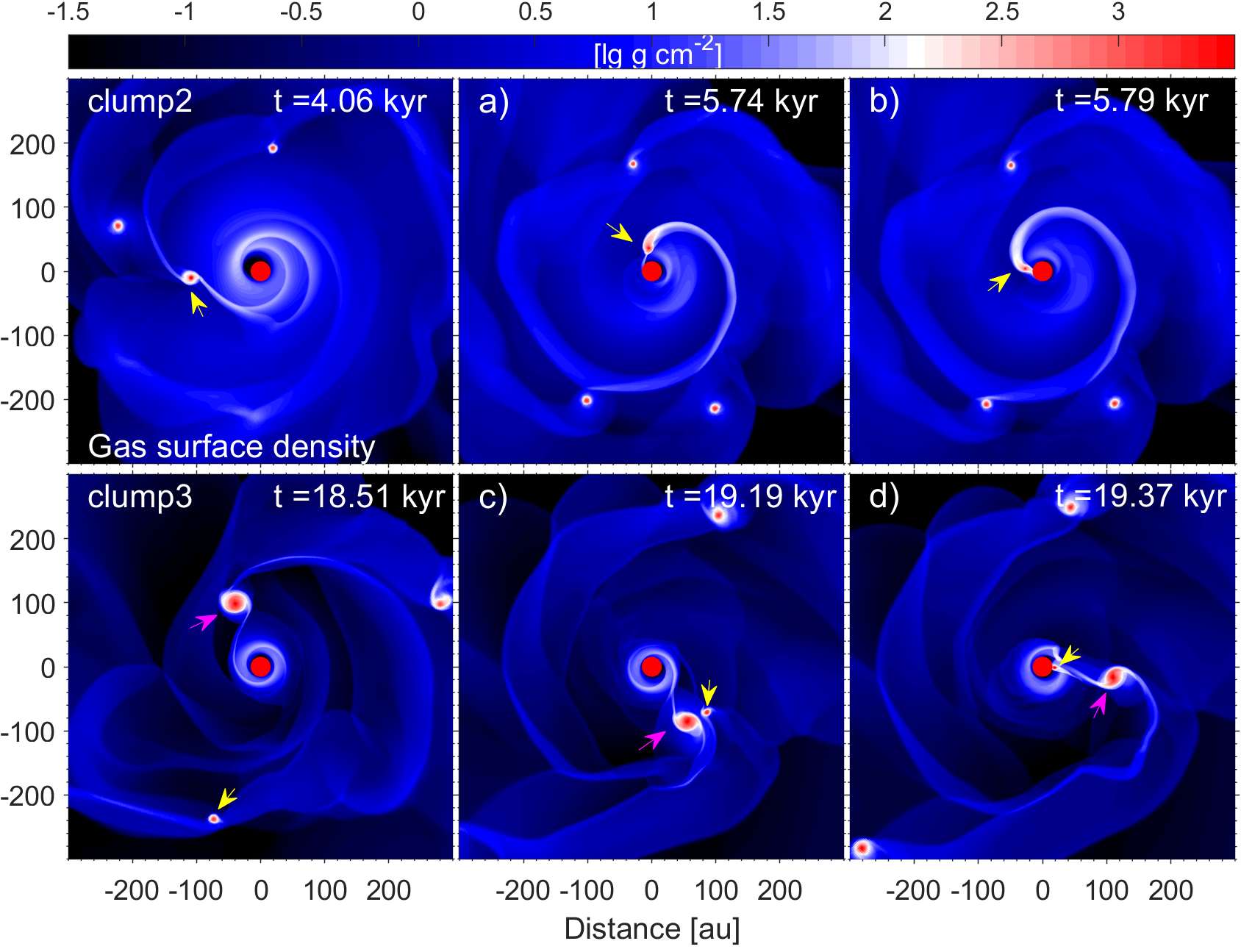}
\par\end{centering}
\caption{\label{fig:6} Gas surface density maps in the inner 600$\times$600~au$^2$ box in the clump-infall model. The top and bottom rows show the disk structure corresponding to the Clump2 and Clump3 bursts in the second and third rows of Fig.~\ref{fig:1}. The first and second columns (from left to right) show the time instances preceding the burst, while the third column presents the disk during the burst.
The yellow arrows show the position of perturbed clumps that fall onto the central star, while the red arrows indicate the clump that is involved in the close encounter.   The disk is rotating counterclockwise.}
\end{figure}

The second type of accretion and luminosity burst is caused by infall of gaseous clumps that form in gravitationally unstable disks via disk fragmentation. This process is illustrated in Figure~\ref{fig:6} showing the gas surface density distributions before and during the burst events that are indicated in the second panel of Figure~\ref{fig:1} with arrows.  Clearly, the disk exhibits a highly fragmented pattern with several clumps and irregular spiral arms. 
As was described in detail in \citet{2018VorobyovElbakyan}, chance encounters between the clumps drive one of them towards the star, while the other is scattered to a wider orbit.  Figure~\ref{fig:6} illustrates two such events. In the top row the encounter is mild and the clump highlighted by the arrow gradually spirals down on the star, leaving  behind a characteristic spiral-like tail. This structure is composed of the clump envelope material that is tidally lost when the clump spirals down towards the star.  The bottom row shows a more dramatic encounter when a small clump is sling-thrown on the star by a massive clump. The falling clump is indicated by the arrow and is approaching the star almost radially. 


\begin{figure}
\begin{centering}
\includegraphics[width=1\columnwidth]{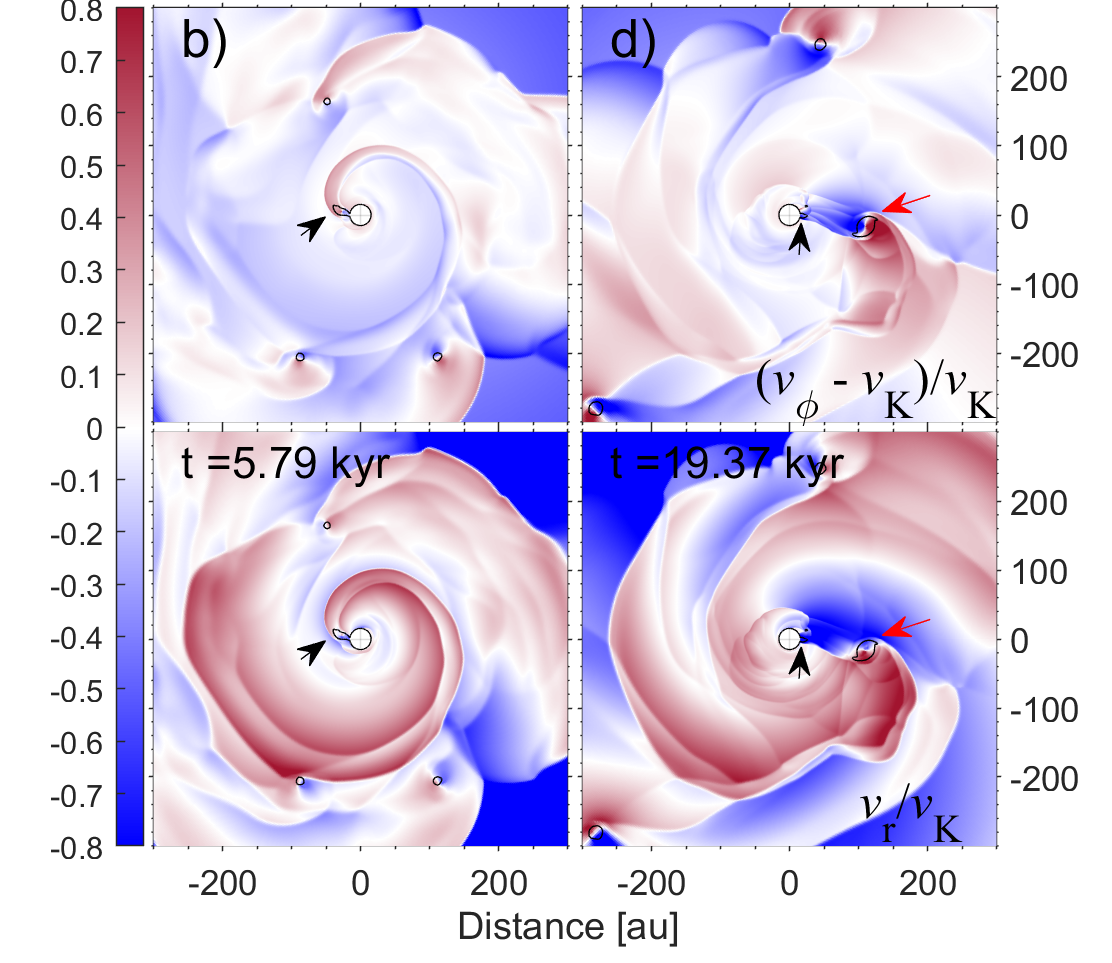}
\par\end{centering}
\caption{\label{fig:8} Spatial distributions of the residual azimuthal velocity $\delta v_\phi=(v_{\phi} - v_{\rm K})/v_{\rm K}$ (top panels) and the ratio of radial to Keplerian velocity  $\delta v_r=v_{\rm r}/v_{\rm K}$ (bottom panels) in the clump-infall model. Left and right columns correspond to the time instances during the Clump2 and Clump3 bursts, which are marked in the right column of Figure~\ref{fig:6} with letters b) and d). The black contour lines outline the clumps, the black arrows point to the position of the clump that causes the burst, while the red arrows indicate the position of the clump that participated in the close encounter (but did not migrate to the star).  The white circles in the coordinate center represent the sink cell. The disk rotates counterclockwise.}
\end{figure}

Figure~\ref{fig:8} displays the spatial maps of the residual azimuthal velocity $\delta v_\phi$ and ratio $\delta v_r$ in the clump-infall models. The left and right columns correspond to panels b) and d) in  the top and bottom rows of Figure~\ref{fig:6}. 
Consider first the left column. Clearly, the in-spiraling clump produces strong perturbations from the purely Keplerian pattern of rotation, reaching tens of per cent in some parts of the disk.  The clump rotates counterclockwise, in the same direction as the disk, and shows a characteristic pattern with a positive $\delta v_\phi$ lying at larger distances and negative $\delta v_\phi$ at smaller distances with respect to the center of the clump ($\approx 25$~au). The trailing spiral arm behind the clump expands outwards as the result of angular momentum exchange with the infalling clump. 
The velocity pattern  shown in the right column of Figure~\ref{fig:8} carries signatures of a violent close encounter, which threw the smaller clump towards the star and triggered the burst. The region between the two clumps is characterized by a prominent inward flow ($\delta v_r <0$). The residual azimuthal velocity $\delta v_\phi$  is also negative in this region. The larger and more distant clump creates a notable expanding wake with positive $\delta v_r$ and $\delta v_\phi$.

\begin{figure}
\begin{centering}
\includegraphics[width=1\columnwidth]{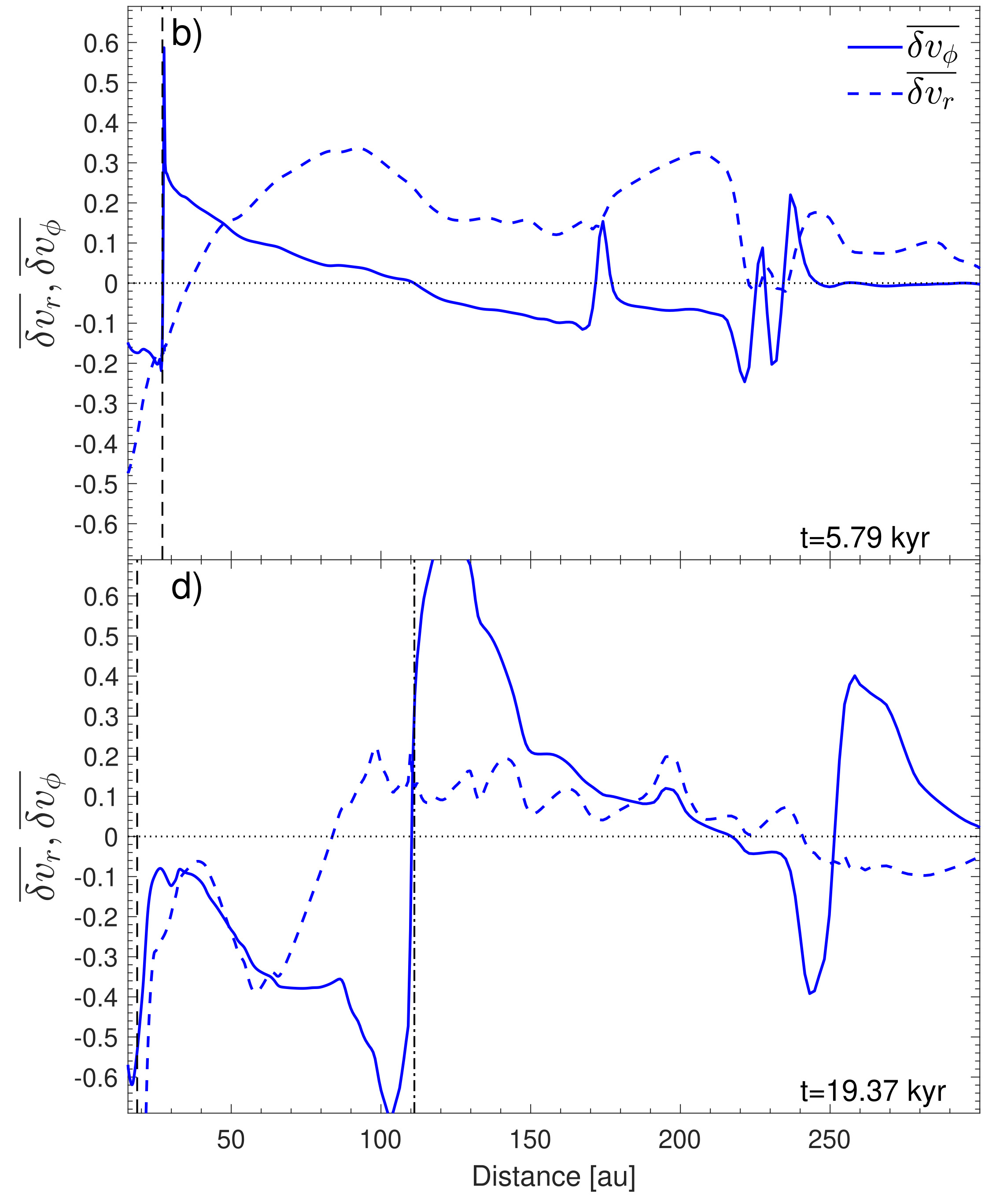}
\par\end{centering}
\caption{\label{fig:9}  Azimuthally averaged radial profiles of residual velocity $\delta v_\phi$ and ratio $\delta v_r$ in the inner 300~au for the clump-infall models. The top and bottom panels correspond to the right and left columns in Figure~\ref{fig:8} (or to the Clump2 and Clump3 bursts, respectively). The vertical dashed lines show the radial distance of the infalling clumps that cause the burst. The dash-dotted line in the bottom panel shows the radial distance of the clump (shown with the red arrow in Figure~\ref{fig:8}) participating in the close encounter that triggered the burst.}
\end{figure}

\begin{figure*}
\begin{centering}
\includegraphics[width=2\columnwidth]{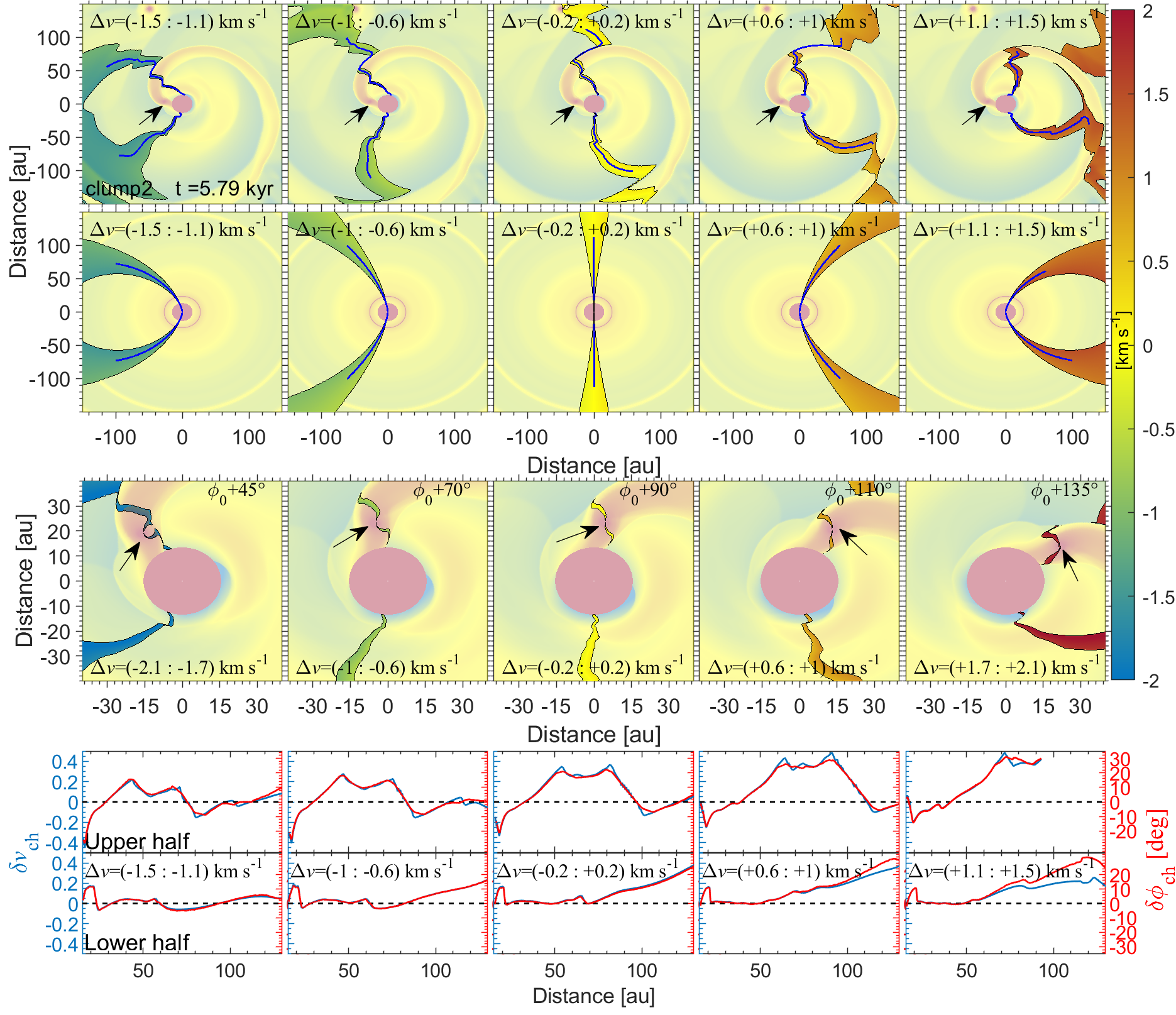}
\par\end{centering}
\caption{\label{fig:17a} Line-of-sight velocity channel maps for the Clump2 burst at its peak luminosity (first row) vs. those  of an idealized Keplerian disk (second row). The black arrows mark the position of the clump causing the burst.  The third row zooms in on the clump, which is viewed at different azimuthal angles as indicated in each panel. The disks are tilted by 30$^\circ$ with respect to the horizontal axis with the upper part being further out from the observer. The color bar presents the deviation from the zero-velocity in km~s$^{-1}$. The velocity intervals are indicated in each panel (note the difference in the zoom-in case).  The corresponding gas surface density maps are plotted in pale palette for convenience. The rotation is counterclockwise. The bottom row presents the deviations $\delta \phi_{\rm ch}$ and $\delta v_{\rm ch}$ for the corresponding velocity channels.  The values for the upper and lower halves of the disk are shown separately. The centers of the model and  Keplerian channels are shown by the blue curves in the top and middle panels for convenience. }
\end{figure*}

To better quantify a strongly perturbed character of the velocity field in the clump-infall model, we plot
the azimuthally averaged residual velocities in Figure~\ref{fig:9}. The top and bottom panels correspond to the left and right columns in Figure~\ref{fig:8} (or to the Clump2 and Clump3 bursts). Clearly, the azimuthally averaged values of $\delta v_\phi$ and $\delta v_r$ show a much stronger deviation amplitude from the purely Keplerian rotation than in the MRI model (see Fig.~\ref{fig:4}). Let us first consider the top panel, which corresponds to the Clump2 burst. The  transition from infalling to expanding motion as indicated by the change in the sign of $\overline{\delta v_r}$ occurs near the position of the in-spiraling clump, which is being disintegrated by tidal torques.  The material interior to the clump flows towards the star causing the burst, while the material behind and further out with respect to the clump position (the spiral wake in Fig.~\ref{fig:6}) is pushed outwards. The gravitational exchange of angular momentum between the clump and the spiral wake creates this velocity pattern. A sharp switch from sub-Keplerian to super-Keplerian motion is also evident at the position of the clump.  The clump is fast rotating counterclockwise and this is reflected in the amplitude of the jump. The spiral arm behind the clump gains angular momentum, accelerates, and expands, contributing to super-Keplerian rotation
in the region between 25 and 100 au. 


Let us now consider the bottom panel in Figure~\ref{fig:9} corresponding to the Clump3 burst. The azimuthally averaged velocity distribution is notably different and shows a dominant inward-flowing, sub-Keplerian pattern in the inner 100~au of the disk. This pattern is caused by the clump that is sling-thrown toward the star during the close encounter with a more massive clump located at $\approx 110$~au (see panel c in Fig.~\ref{fig:6}). On the contrary, the regions outside 110~au  demonstrate a super-Keplerian expansion as the result of gravitational exchange of angular momentum between the two clumps during the close encounter. 
Velocity perturbations at distances beyond 150~au are caused by other distant clumps that can be seen in Figure~\ref{fig:6}.
Overall, the clump-infall models are characterized by a peculiar velocity pattern with a much stronger deviation from Keplerian rotation than the MRI models.

\begin{figure}
\begin{centering}
\includegraphics[width=1\columnwidth]{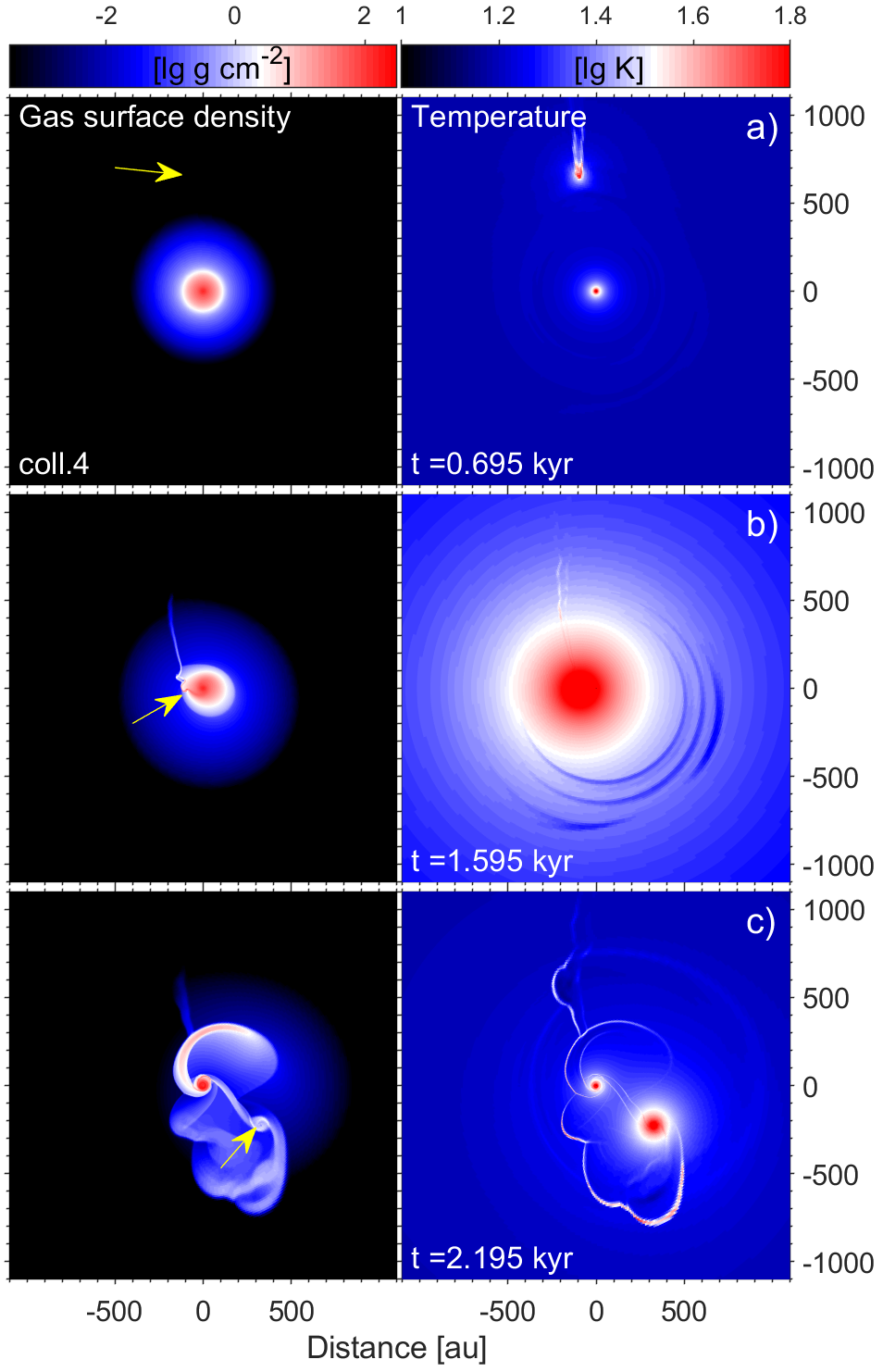}
\par\end{centering}
\caption{\label{fig:10} Spatial distributions of gas surface density (left column) and temperature (right column) during the close encounter that triggers the Collision4 burst (see Fig.~\ref{fig:1}). The arrow points to the intruder on a prograte trajectory. The time is counted from the launch of the intruder. The scale bars are in log g~cm$^{-2}$ and log K.   }
\end{figure}

\begin{figure}
\begin{centering}
\includegraphics[width=1\columnwidth]{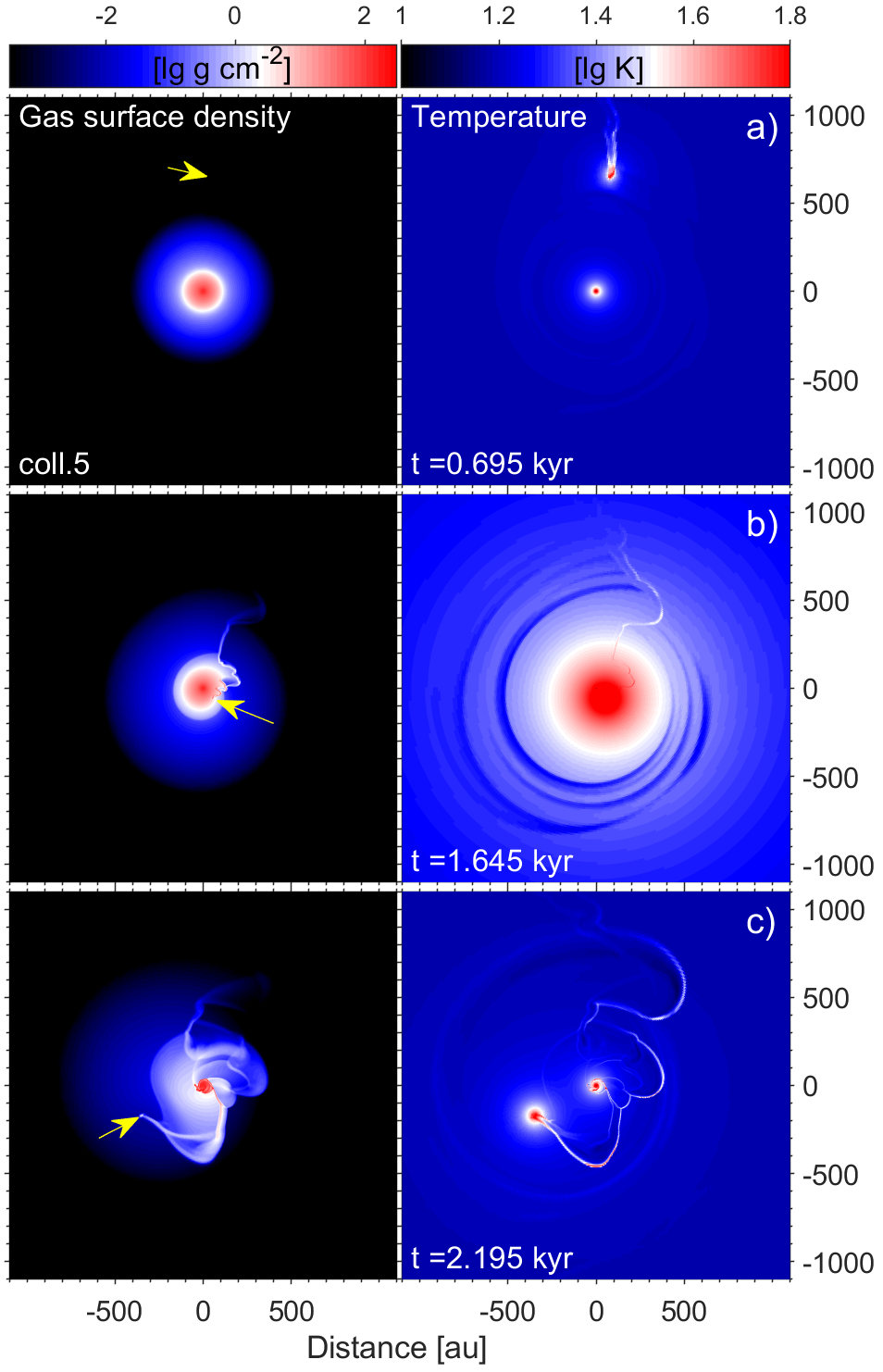}
\par\end{centering}
\caption{\label{fig:11} Similar to Fig.~\ref{fig:10} but for the retrograde collision that triggers the Collision5 burst.}
\end{figure}

The comparison of velocity channel maps for the Clump2 burst at its maximum luminosity ($t=5.79$~kyr) with those of an idealized Keplerian disk is shown in Figure~\ref{fig:17a} and confirms our previous findings.  The model and Keplerian disks are both tilted at 30$^\circ$ with respect to the horizontal axis with the upper part being further out from the observer. The black arrow shows the position of the clump causing the burst. Clearly, strong perturbations in the gas surface density and velocity field (see Fig.~\ref{fig:9}) caused by the clump also lead to strong deviations of the velocity channel maps from the Keplerian ones throughout the entire disk. The kinks in the velocities for the Clump2 burst are more pronounced compared to those for the MRI1 burst. Moreover, the velocity channels are often discontinuous at the leading and trailing edges of the spiral arc that is formed behind the in-spiraling clump, indication strong non-Keplerian motion caused by rotation-expansion motion of the arc.
We conclude that velocity channel maps can be used to infer strong perturbations caused by the in-spiraling clump. As for the clump itself, the situation is more complex and depends on the azimuthal angle at which the system is viewed. The inner regions where the clump is being tidally disintegrated may also be difficult to resolve. The third row of Figure~\ref{fig:17a} zooms in on the clump at different azimuthal angles, where $\phi_0$ is the azimuthal angle of the clump in the top row. Clearly, at certain disk orientations the velocity channel maps in the vicinity of the clump are strongly distorted, likely due to strong internal rotation of the clump. However, there are orientations at which the distortion is minimal.

To provide a quantitative analysis for the velocity channel maps during the Clump2 burst, we show in the bottom row of Figure~\ref{fig:17a} the angle and velocity deviations of the model velocity channels from those of an idealized Keplerian disk ($\delta \phi_{\rm ch}$ and $\delta v_{\rm ch}$, respectively) calculated using Equations~(\ref{angle_ch}) and (\ref{vel_ch}).
These deviations quantify the degree of distortion of the corresponding channel maps in the physical and velocity space with respect to the channel maps of an idealized Keplerian disk.
The deviations are clearly much stronger than those of the MRI model (see Fig.~\ref{fig:16a}). For instance, the mismatch between the centers of the model and Keplerian velocity channels ($\delta \phi_{\rm ch}$) can amount to tens of degrees and the corresponding relative deviation in the line-of-sight velocities $(\delta v_{\rm ch})$ can be on the order of tens of percent. The radial profiles of $\delta \phi_{\rm ch}$ and $\delta v_{\rm ch}$ are distinct for the upper and lower halves of the disk experiencing a clump-triggered burst. This is in stark contrast to the corresponding profiles in the MRI-burst model, which tend to follow a similar pattern, albeit with notable variations. The distinct character of the deviations in the upper and lower halves of the disk in the clump-infall model is the result of strong asymmetry that is typical of the disk with an in-spiraling clump. 

Finally, we note that strong deviations from the Keplerian rotation pattern are typical not only for the time instances during the clump-triggered bursts, but also for a gravitationally fragmented disk in general. For instance, the characteristic jumps in the azimutal velocity deviations ($\overline{\delta v_\phi}$) at the position of the clumps indicated with the vertical dash-dotted lines in Figure~\ref{fig:9} are also found at the time instances preceding the burst ($t=4.06$~kyr). Distortions in the model channel maps of a fragmented disk as compared to the symmetric Keplerian pattern are also appreciably stronger than what was found for a merely gravitationally unstable (but fragmentationally stable) disk. Strong deviations from the Keplerian rotation in the absence of an obvious intruder star (see Sect.~\ref{Collisions}) may signalize that the system is prone to the accretion burst activity.

\subsection{Bursts triggered by collisions} 
\label{Collisions}
The last type of accretion burst considered in this work is triggered by close encounters of an intruder star with a target disk. Figures~\ref{fig:10} and \ref{fig:11} display the gas surface density and temperature distributions during close encounters that trigger the Collision4 and Collision5 bursts (see the fourth and fifth panels in Fig.~\ref{fig:1}). The initial intruder mass is 0.5~$M_\odot$, which is approximately equal to the mass of the target star (see Table~\ref{tab:1}). The intruder was set on a collision trajectory at 1500~au from the target disk, which guarantees a smooth initial start. Two types of collision are considered: the prograde one (Figure~\ref{fig:10}) and retrograde one (Figure~\ref{fig:11}). From the calculated eccentricity of target's trajectory ($e>1$) we concluded that the collisions are hyperbolic,  which is a consequence of the chosen initial conditions for the intruder.

The first row in both figures shows the time instance when the intruder approaches the disk. The ambient gas density is too low so that the intruder is only seen in the temperature distribution through heating of the surrounding medium. The second row corresponds to the time instance of closest approach between the target and intruder stars. The periastron distances are 82.4~au (Figure~\ref{fig:10}) and 75.3~au (Figure~\ref{fig:11}), while the corresponding velocities of the intruder are 5.21 and 5.43~km~s$^{-1}$. We experimented with different periastron distances and found that collisions with $r_{\rm per}>150$~au produce weak outbursts hardly exceeding 30~$L_\odot$. { As the periastron distance increases, the intruder produces less gravitational perturbation to the disk of the target star, so that the intruder still dominates in terms of the burst luminosity of the intruder-target system. For as long as we consider a disk-penetrating encounter, the majority of the brightening is therefore due to mass accumulated by the intruder, but we are interested in the observable kinematic signatures on the target star's disk.}
Collisions with $r_{\rm per}\la$~a~few~$\times 10$~au lead to numerical instabilities and cannot be completed. Such close encounters should, however, be quite rare \citep{2010ForganRice}. During the closest approach the disk is already notably perturbed but the strongest response is seen in the temperature distribution, which indicates a strong heating event caused by the burst.
The third row presents the time instance when the intruder starts receding. At this stage, the disk is strongly perturbed and exhibits characteristic spiral-like tails \citep[see][for details on these structures]{2020VorobyovTails}. The intruder is still brighter than the target in the temperature distribution. Strong shock waves cased by the intruder passage are also evident in the gas temperature. 

\begin{figure}
\begin{centering}
\includegraphics[width=1\columnwidth]{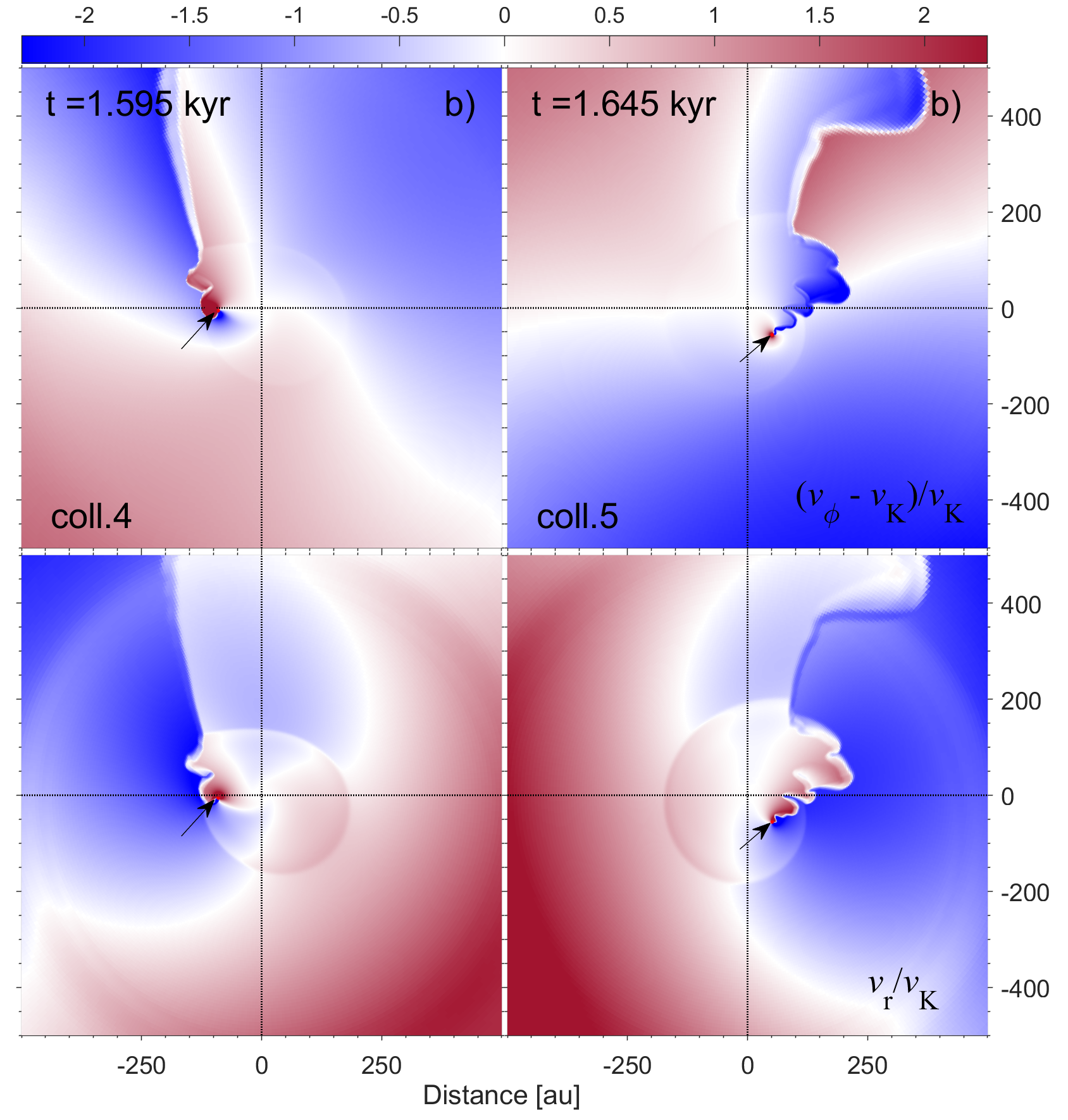}
\par\end{centering}
\caption{\label{fig:13} Spatial distributions of the residual azimuthal velocity $\delta v_\phi=(v_{\phi} - v_{\rm K})/v_{\rm K}$ (top panels) and the ratio of radial to Keplerian velocity  $\delta v_r=v_{\rm r}/v_{\rm K}$ (bottom panels) in the collision models. The left and right columns correspond to the peaks of Collision4 and Collision5 bursts (see the second row in Figs.~\ref{fig:10} and \ref{fig:11}). The black arrows point to the position of the intruder.  The intersection of the dashed black lines marks the position of the target star. The disk rotates counterclockwise.}
\end{figure}

The spatial maps of residual velocities $\delta v_\phi$ and $\delta v_r$ at the peak of the Collision4 and Collision5 bursts are displayed in Figure~\ref{fig:13}. We emphasize that all velocities in the collision models are provided in the local frame of reference of the target star, which also moves in response to the gravitational force of the intruder (the simulations are performed in the non-inertial frame of reference of the target). As can be expected, strong deviations from the Keplerian rotation of the target disk are present. A certain reflection symmetry is also notable in the residual velocities of prograde (Collision4) and retrograde (Collision5) models. This is most clearly seen for $\delta v_r$ but is also evident in the lower region for $\delta v_\phi$. 

\begin{figure}
\begin{centering}
\includegraphics[width=1\columnwidth]{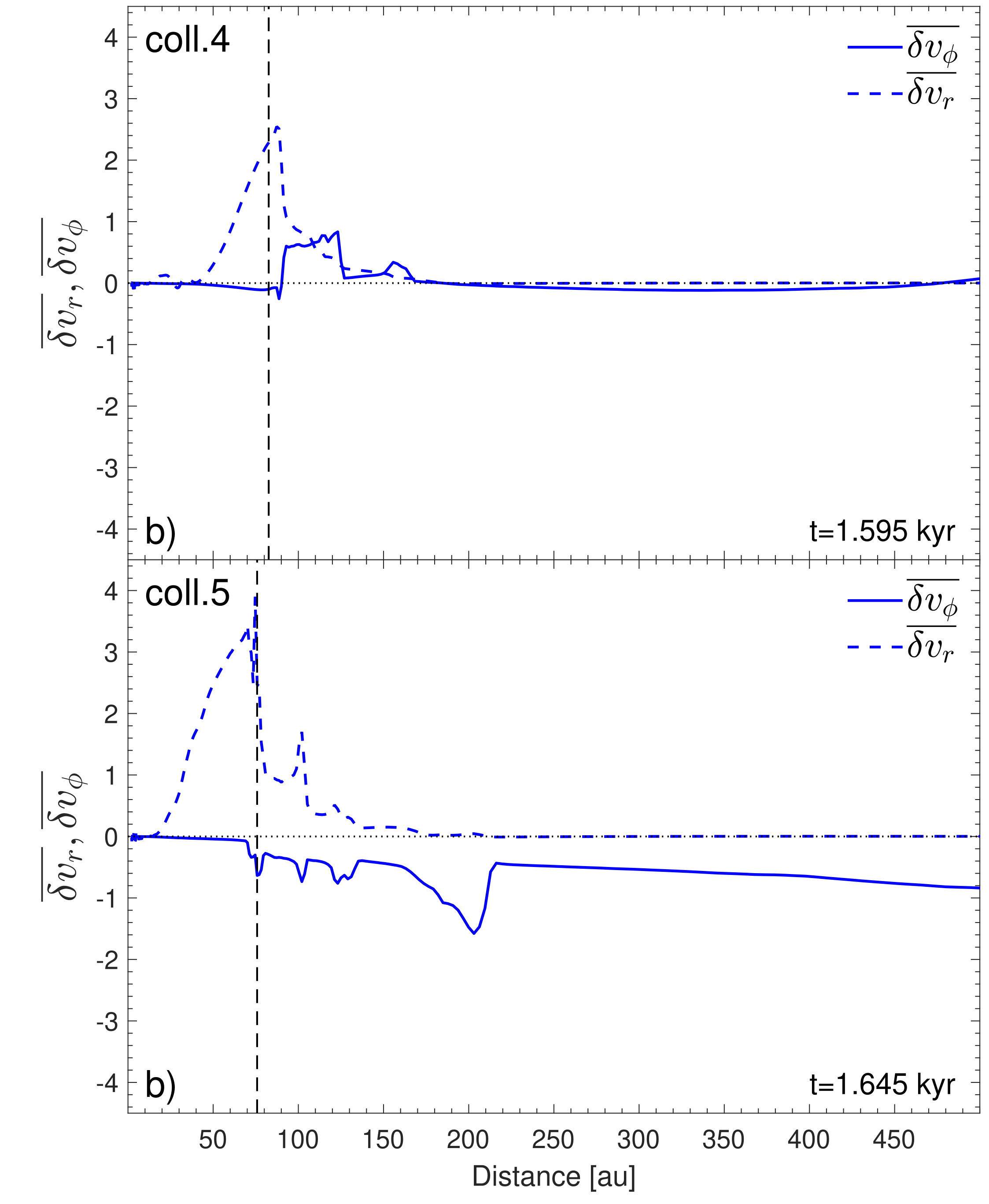}
\par\end{centering}
\caption{\label{fig:14} Azimuthally averaged radial profiles of residual velocity $\delta v_\phi$ (solid line) and ratio $\delta v_r$ (dashed line) in the inner 500~au for the Collision4 (top) and Collision5 (bottom) bursts. The top and bottom panels correspond to the left and right columns in Figure~\ref{fig:13}. The vertical dashed lines show the radial distance of the intruder.}
\end{figure}

Figure~\ref{fig:14} presents the azimuthally averaged residual velocities $\overline{\delta v_\phi}$ and $\overline{\delta v_r}$ for the prograde and retrograde collision models at the peak of the burst. A comparison with the corresponding Figures~\ref{fig:4} and \ref{fig:9} for the MRI and clump-infall bursts reveals that the collision bursts are characterized by the strongest perturbations to the Keplerian velocity field.  The perturbations within 150--200~au can be a factor of several stronger than the Keplerian velocity of the target disk. The radial flows are strongest inside and in the vicinity of the intruder position (see the vertical dashed lines), while the azimuthal velocity is notably perturbed at radial distances beyond the intruder position.

\begin{figure*}
\begin{centering}
\includegraphics[width=2\columnwidth]{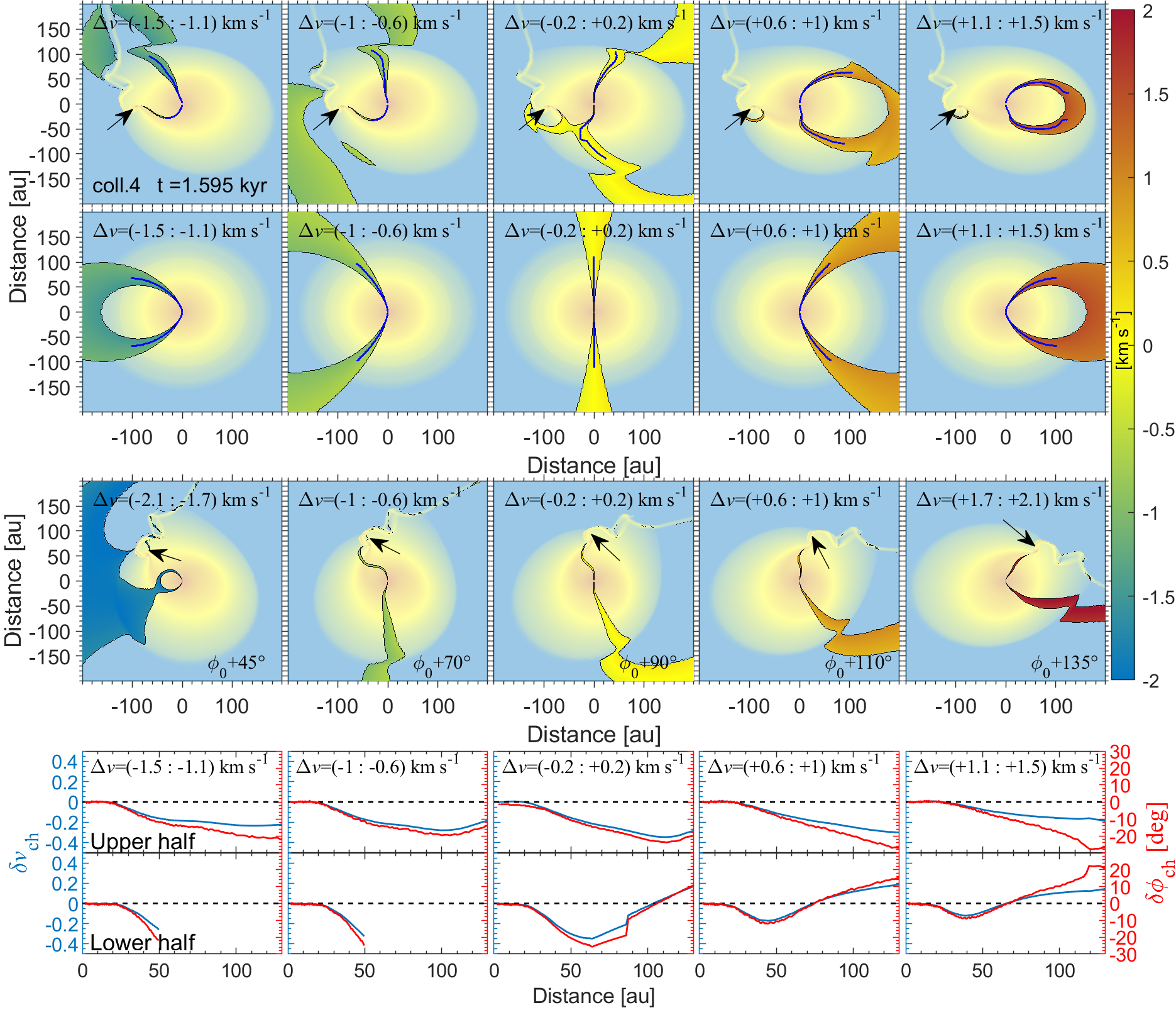}
\par\end{centering}
\caption{\label{fig:18} Line-of-sight velocity channel maps for the Collision4 burst at its peak luminosity (first row) vs. those  of an idealized Keplerian disk (second row). The black arrows mark the position of the intruder that triggers the burst. The third row zooms in on the intruder, which is viewed at different azimuthal angles as indicated in each panel. The disks are tilted by 30$^\circ$ with respect to the horizontal axis with the upper part being further out from the observer. The color bar presents the deviation from the zero-velocity in km~s$^{-1}$. The velocity intervals are indicated in each panel (note the difference in the zoom-in case).  The corresponding gas surface density maps are plotted in pale palette for convenience. The rotation is counterclockwise. The bottom row presents the deviations $\delta \phi_{\rm ch}$ and $\delta v_{\rm ch}$ for the corresponding velocity channels.  The values for the upper and lower halves of the disk are shown separately. The centers of the model and  Keplerian channels are shown by the blue curves in the top and middle panels for convenience. } 
\end{figure*}

Figure~\ref{fig:18} presents the comparison of velocity channel maps for the Collision4 burst at its luminosity peak ($t=1.595$~kyr) with the channel maps of an idealized Keplerian disk of the target star. Both disks are inclined by $30^\circ$ with respect to the horizontal axis with the upper part being further out from the observer. This type of the burst is characterized by strongest deviation from Keplerian rotation, as was already noted before. The kinks in velocities are now located  at the disk edges, indicating the expansion and distortion of the target disk in response to the encounter event. The velocities are also strongly distorted from the Keplerian pattern in the vicinity of the intruder. The velocity gradients become much stronger there, which is reflected in the velocity channel maps becoming narrower near the intruder. The third row shows the system at different azimuthal angles, where $\phi_0$ is the azimuthal angle of the intruder in the top row. The picture is qualitatively similar when the system is viewed from different azimuthal angles -- the channel maps near the intruder are extremely narrow and patchy, reflecting a highly perturbed velocity environment.

Finally, the bottom row in Figure~\ref{fig:18} presents the angle and velocity deviations $\delta \phi_{\rm ch}$ and $\delta v_{\rm ch}$ calculated using Equations~(\ref{angle_ch}) and (\ref{vel_ch}).  The values of $\delta \phi_{\rm ch}$ and $\delta v_{\rm ch}$ are of a similar magnitude when compared to those in the clump-infall model (see Fig.~\ref{fig:17a}), but are much stronger than in the MRI-model (see Fig.~\ref{fig:16a}). Nevertheless, in some velocity channels (the two rows on the left) the velocity pattern of the target disk is so strongly distorted that we have difficulty in calculating the corresponding values of $\delta \phi_{\rm ch}$ and $\delta v_{\rm ch}$ beyond 50 au.  Moreover, the radial profiles of $\delta \phi_{\rm ch}$ and $\delta v_{\rm ch}$ have a pattern that is distinct form those of the clump-infall and MRI-triggered bursts. The pattern is notably asymmetric with respect to the upper and lower halves of the disk only on the blue-shifted part of the disk where the intruder is located. The channel maps approach a mirror symmetry on the opposite side from the intruder (red-shifted velocity channels), although strong deviations from the Keplerian rotation are still present. Overall, negative deviations dominate, reflecting the distortion of the target disk owing to the gravitational pool of the intruder star during the prograde encounter. We note that strong deviations from the Keplerian rotation linger in the disk of the target for a time period that is longer than the burst duration. However, the character of these deviations changes with time. In particular, the mirror symmetry on the opposite side from the intruder disappears. We plan to study in more detail the long-term evolution of the channel maps in follow-up studies.

\section{Model caveats}
\label{Caveats}
In this section, we briefly review several model caveats that are associated with our study.
In the MRI model, the peak $\alpha$-value in the disk regions involved in the burst ($\alpha_{\rm max}$) is found to affect notably the burst appearance.  
In the $\alpha_{\rm max}=1.0$ case, the bursts are the sharpest and strongest, while in the $\alpha_{\rm max}=0.01$ the bursts are of lower amplitude, longer duration, and lesser frequency. This is not unexpected, since the $\alpha$-parameter enters the coefficient of kinematic viscosity, which in turn determines the characteristic viscous time of the problem and the burst duration. The value of $\alpha$-parameter is found by calculating the Maxwell stress tensor in numerical simulations of MHD turbulence and varies in wide limits. While $\alpha=1.0$ is a theoretical upper limit and is unlikely to be reached in real circumstances, values in the 0.01--0.1 limits are possible \citep[e.g.,][]{2018Yang,2020Zhu}, may depend on disk conditions and magnetic field strength, and may vary from burst to burst.  Further MHD studies of this phenomenon are needed to better constrain the value of $\alpha$ if possible.

In the clump-infall model the burst characteristics may depend on the placement of the inner disk boundary. In the present study, we set it at 15~au for the clump-infall models to achieve a high resolution on the log-spaced numerical grid in the $r$-direction. The smallest inner boundary of 5~au was set in the clump-infall models of \citet{2015VorobyovBasu}, which also showed luminosity bursts. Nevertheless, the  physical mechanisms that operate in the innermost regions may alter the light curve but are unlikely to change the global disk kinematics that is characteristic of clump-triggered bursts.

For the cases with clump-infall or collision-triggered bursts, the systematic analysis of the observed gas velocity profiles remains challenging.
When the disk morphology is not axisymmetric, it may not be possible to measure the inclination of the disk. Therefore, how to de-project the resolved gas structures in the line-of-sight dimension is ambiguous. Owing to this uncertainty,  it may be difficult to exactly measure how the velocities of the resolved gas structures are deviated from the Keplerian rotation curve.

Finally, the collision-triggered bursts were simulated in a simplified in-plane geometry and with a parametric rather than hydrodynamic calculation of accretion rates on the intruder. According to \citet{2010ForganRice}, highly inclined collisions do not produce strong bursts. A moderate inclination may, nevertheless, affect the burst duration, its shape, and kinematic signatures. The chosen parametric method for calculating the mass accretion rate on the intruder star may also affect our results. This is a fundamental problem in the grid-based numerical codes, such as FEOSAD, and we based our choice on previous investigations of, e.g.,  \citet{2010FederrathBanerjee}. We did not explore a possibility that the intruder star may also host a disk. Such a case was considered in the recent smoothed-particle hydrodynamics simulations of \citet{2020Cuello} and showed accretion bursts of a much smaller amplitude ($\sim 10^{-6} M_\odot$~yr$^{-1}$ during the peak) than is expected for most FUors.

\section{Conclusions}
\label{Conclude}
We compared the accretion and luminosity bursts produced by three distinct mechanisms: MRI in the innermost disk regions, infall of clumps formed via disk gravitational fragmentation, and close encounter with an intruder star. For this purpose, we used the numerical hydrodynamics code FEOSAD, which computes the disk formation and dynamics in the thin-disk limit.  The MRI bursts were modelled using the adaptive $\alpha$-approach based on the layered disk model \citep{2001Armitage, 2014BaeHartmann, 2019Kadam}, the clump-infall bursts were modelled by simulating the dynamics of a strongly gravitationally unstable disk \citep{2018VorobyovElbakyan}, while the collision bursts were triggered by sending an intruder star of $0.5~M_\odot$ on a collision trajectory with a protoplanetary disk. We investigated in detail the  gas velocity fields and velocity channel maps for each burst mechanism. 
Our conclusions can be summarized as follows.

-- The considered burst mechanisms operate in circumstellar disks that are characterized by peculiar kinematic features, which in turn may help to distinguish between different burst origins. The disks in the collision and clump-infall models are characterized by strong gas velocity deviations from the Keplerian rotation (reaching tens of per cent for the radial and azimuthal components), while the disks in the MRI models are  characterized only by mild deviations (a few per cent) that are mostly caused by the gravitational instability that fuels the bursts. 

-- The velocity channel maps reflect the kinematic perturbations of the disk in each considered burst model. For the MRI bursts, the velocities are characterized by kinks and wiggles that are typical for gravitationally unstable disks \citep{2020Hall}. For the clump-infall models, the kinks are stronger and are even discontinuous at the edges of expanding spiral arcs created by in-spiraling clumps. For the collision models, the kinks are strongest and discontinuous near the edges of the target disk. The channel maps are extremely narrow in the vicinity of the intruder, thus indication strong velocity gradients and a highly perturbed velocity environment.

-- The deviations of the model velocity channels from a symmetric pattern that is expected for a purely Keplerian disk are distinct among the considered burst models, which may be helpful when distinguishing between the burst mechanisms. In particular, these deviations are strongest in the clump-infall and collision models and show notable asymmetries with respect to the upper and lower halves of the disk, while in the MRI model the deviations are weaker and posses a higher degree of symmetry.

-- Among the three burst mechanisms, the MRI bursts are less energetic in terms of the peak luminosity during the burst, although the strength of the collision bursts depends sensitively on the periastron distance during the closest approach.  
The burst amplitudes vary in the $\Delta m =2.5-3.7$ limits, except for the clump-infall model where $\Delta m$ can reach 5.4. We note that a small considered sample (from a much larger parameter space) and  boundary conditions may affect the derived amplitudes.

{It is worth noting that the disk spatial structures in the considered models are very different from one another and may provide important clues for interpreting the disk kinematics. Therefore, the combination of velocity information and continuum imaging should complement one another and aid in understanding the burst mechanisms. We warn, however, that continuum imaging traces grown dust, which spatial distribution may be different from that of gas. Numerical simulations that take dust dynamics and growth into account \citep[e.g.,][]{2020VorobyovElbakyanTakami,2020VorobyovKhaibrakhmanov} are needed to better understand the burst mechanisms.}
Future studies in this direction should also include the construction of synthetic velocity channel maps in the emission lines of the disk tracer species (such as CO isotopologues), as well as a comparative analysis of the model light curves for different burst-triggering mechanisms, preferably using similar numerical setups and/or codes.

\section*{Acknowledgements}
We are thankful to the anonymous referee for useful suggestions that helped to improve the manuscript. This work was supported by the Russian Fund for Fundamental Research, Russian-Taiwanese project 19-52-52011 and MoST project 108-2923-M-001-006-MY3. H.B.L. is supported by the Ministry of Science and
Technology (MoST) of Taiwan, grant No. 108-2112-M-001-002-MY3. The simulations were performed on the Vienna Scientific Cluster.

\begin{appendix}

\section{Observational characteristics of known FUors}
For the reader convenience, Table~\ref{tab:3} provides
a compilation of the main characteristics of FUors and FUor-like objects (the latter are named in the sense that their onset time is unknown).  The vertical dashed line separates these two sub-classes (with FUors going first). The luminosity $L_{\rm obs}$ refers to the maximum value detected during the provided observation period, and $t_{\rm bst}$ and $t_{\rm rise}$ are the burst durations and rise times, respectively.

\begin{table}
\label{FUors}
\center
\caption{\label{tab:3}Observational characteristics of FU Orionis objects}
\resizebox{\columnwidth}{!}{\begin{tabular}{ccccccccc}
\hline 
\hline 
Object  & $L_{\mathrm{obs}}$ &  $t_{\mathrm{bst}}$ & $t_{\mathrm{rise}}$ & Distance  & Period of & Ref. \tabularnewline
  & [$L_{\odot}$] & [yr] & [yr] & [pc] &  observ. &  \tabularnewline
\hline 
RNO1B/RNO1C     & 1600$\dagger$		&$>$100	  & 0-12    & 930 	& 2006-2010 	& 1,2,3 \tabularnewline
SVS13A          & 25		&$>$30	  & 2   	& 235 	& - 	& 4 \tabularnewline
Gaia 17bpi      & 7.5		&$>$4	  & 1 	    & 1270 	& - 	& 7 \tabularnewline
V2775 Ori       & 29		& 11-21	  & 0-10 	& 428 	& 2006-2010 	& 3,8 \tabularnewline
FU Ori          & 320		&$>$84	  & 1 	    & 416 	& 2004 	& 9,10 \tabularnewline
V900 Mon        & 184		&$>$22	  & 1-2 	& 1500 	& 2006-2010 	& 3 \tabularnewline
V960 Mon        & 106		& 6	  	  & 1 	    & 1638 	& 2006-2010 	& 3,13 \tabularnewline
V1515 Cyg       & 180		&$>$70	  & 10  	& 1009 	& 2004 	& 9,10 \tabularnewline
HBC722          & 20		&$>$10	  &$<$0.2 	& 771 	& 2016 	& 14,15 \tabularnewline
V1057 Cyg       & 400		& 30-40	  & 1 	    & 920 	& 2004 	& 9,10 \tabularnewline
V2494 Cyg       & 129		&$>$31	  & 0-6	    & 498 	& 2006-2010 	& 3,16 \tabularnewline
V2495 Cyg       & 21		&$>$19	  & 0-1 	& 600 	& 2006-2010 	& 3,17 \tabularnewline
V1735 Cyg       & 72		& 60	  & 0-8 	& 624 	& 2006-2010 	& 3,18 \tabularnewline
V733 Cep        & 30		&$>$67	  & 0-18 	& 669 	& 2006-2020 	& 18-20 \tabularnewline
\hdashline
PP13S           & 51		&$>$100	  & -   	& 450 	& 2006-2010 	& 3,5 \tabularnewline
L1551 IRS5      & 29$\dagger$		&$>$100	  & -  	    & 147   & 2006-2010 	& 3,6 \tabularnewline
Haro 5a/6a IRS  & 18$\dagger$			&$>$100	  & - 	    & 388 	& 2006-2010 	& 3 \tabularnewline
V883 Ori        & 103		&$>$100	  & - 	    & 388 	& 2006-2010 	& 3 \tabularnewline
NGC2071 MM3     & 35		&$>$100	  & - 	    & 388 	& 2006-2010 	& 3 \tabularnewline
AR 6A/6B        & 310$\dagger$			&$>$100	  & - 	    & 738 	& 2006-2010 	& 3 \tabularnewline
Parsamian 21    & 16		&$>$100	  & - 	    & 500 	& 2006-2010 	& 3 \tabularnewline
BBW 76          & 106		&$>$120	  & - 	    & 1093 	& 2004 	& 10,11 \tabularnewline

\hline 
\end{tabular}}
\textbf{Notes.} 1=\citet{1991A&A...244L..13S} 2=\citet{2014ApJ...783..130R} 3=\citet{2018ApJ...861..145C} 4=\citet{2017A&A...599A.121D} 5=\citet{2009ApJ...703...52L} 6=\citet{2007ApJ...671..546L} 7=\citet{2018ApJ...869..146H} 8=\citet{2011A&A...526L...1C} 9=\citet{1977ApJ...217..693H} 10=\citet{2006ApJ...648.1099G} 11=\citet{2013ApJ...778..116N} 12=\citet{2012ApJ...748L...5R} 13=\citet{2015A&A...582L..12H} 14=\citet{2016A&A...596A..52K} 15=\citet{2010A&A...523L...3S} 16=\citet{2013MNRAS.432.2685M} 17=\citet{2003A&A...412..147M} 18=\citet{Peneva} 19=\citet{2007AJ....133.1000R} 20=\citet{Semkov}
$\dagger$=Total luminosity of binary system
\end{table}

\end{appendix}

\bibliographystyle{aa}
\bibliography{Vorobyov}

\newcommand{\noop}[1]{}
\begin{thebibliography}{68}
\expandafter\ifx\csname natexlab\endcsname\relax\def\natexlab#1{#1}\fi

\bibitem[{{Armitage} {et~al.}(2001){Armitage}, {Livio}, \&
  {Pringle}}]{2001Armitage}
{Armitage}, P.~J., {Livio}, M., \& {Pringle}, J.~E. 2001, MNRAS, 324, 705

\bibitem[{{Audard} {et~al.}(2014){Audard}, {{\'A}brah{\'a}m}, {Dunham},
  {Green}, {Grosso}, {Hamaguchi}, {Kastner}, {K{\'o}sp{\'a}l}, {Lodato}, \&
  {Romanova}}]{2014AudardAbraham}
{Audard}, M., {{\'A}brah{\'a}m}, P., {Dunham}, M.~M., {et~al.} 2014, in
  Protostars and Planets VI, ed. H.~{Beuther}, R.~S. {Klessen}, C.~P.
  {Dullemond}, \& T.~{Henning}, 387

\bibitem[{{Bae} {et~al.}(2014){Bae}, {Hartmann}, {Zhu}, \&
  {Nelson}}]{2014BaeHartmann}
{Bae}, J., {Hartmann}, L., {Zhu}, Z., \& {Nelson}, R.~P. 2014, \apj, 795, 61

\bibitem[{{Baraffe} {et~al.}(2017){Baraffe}, {Elbakyan}, {Vorobyov}, \&
  {Chabrier}}]{2017BaraffeElbakyan}
{Baraffe}, I., {Elbakyan}, V.~G., {Vorobyov}, E.~I., \& {Chabrier}, G. 2017,
  \aap, 597, A19

\bibitem[{{Basu}(1997)}]{1997Basu}
{Basu}, S. 1997, ApJ, 485, 240

\bibitem[{{Beck} \& {Aspin}(2012)}]{2012BeckAspin}
{Beck}, T.~L. \& {Aspin}, C. 2012, \aj, 143, 55

\bibitem[{{Caratti o Garatti} {et~al.}(2011){Caratti o Garatti}, {Garcia
  Lopez}, {Scholz}, {Giannini}, {Eisl{\"o}ffel}, {Nisini}, {Massi},
  {Antoniucci}, \& {Ray}}]{2011A&A...526L...1C}
{Caratti o Garatti}, A., {Garcia Lopez}, R., {Scholz}, A., {et~al.} 2011, \aap,
  526, L1

\bibitem[{{Cieza} {et~al.}(2018){Cieza}, {Ru{\'{\i}}z-Rodr{\'{\i}}guez},
  {Perez}, {Casassus}, {Williams}, {Zurlo}, {Principe}, {Hales}, {Prieto},
  {Tobin}, {Zhu}, \& {Marino}}]{2018Cieza}
{Cieza}, L.~A., {Ru{\'{\i}}z-Rodr{\'{\i}}guez}, D., {Perez}, S., {et~al.} 2018,
  \mnras, 474, 4347

\bibitem[{{Connelley} \&
  {Reipurth}(2018{\natexlab{a}})}]{2018ConnelleyReipurth}
{Connelley}, M.~S. \& {Reipurth}, B. 2018{\natexlab{a}}, \apj, 861, 145

\bibitem[{{Connelley} \& {Reipurth}(2018{\natexlab{b}})}]{2018ApJ...861..145C}
{Connelley}, M.~S. \& {Reipurth}, B. 2018{\natexlab{b}}, \apj, 861, 145

\bibitem[{{Contreras Pe{\~n}a} {et~al.}(2017){Contreras Pe{\~n}a}, {Lucas},
  {Minniti}, {Kurtev}, {Stimson}, {Navarro Molina}, {Borissova}, {Kumar},
  {Thompson}, {Gledhill}, {Terzi}, {Froebrich}, \& {Caratti o
  Garatti}}]{2017ContrerasPena}
{Contreras Pe{\~n}a}, C., {Lucas}, P.~W., {Minniti}, D., {et~al.} 2017, \mnras,
  465, 3011

\bibitem[{{Cuello} {et~al.}(2020){Cuello}, {Louvet}, {Mentiplay}, {Pinte},
  {Price}, {Winter}, {Nealon}, {M{\'e}nard}, {Lodato}, {Dipierro},
  {Christiaens}, {Montesinos}, {Cuadra}, {Laibe}, {Cieza}, {Dong}, \&
  {Alexander}}]{2020Cuello}
{Cuello}, N., {Louvet}, F., {Mentiplay}, D., {et~al.} 2020, \mnras, 491, 504

\bibitem[{{De Simone} {et~al.}(2017){De Simone}, {Codella}, {Testi},
  {Belloche}, {Maury}, {Anderl}, {Andr{\'e}}, {Maret}, \&
  {Podio}}]{2017A&A...599A.121D}
{De Simone}, M., {Codella}, C., {Testi}, L., {et~al.} 2017, \aap, 599, A121

\bibitem[{{Dong} {et~al.}(2016){Dong}, {Vorobyov}, {Pavlyuchenkov}, {Chiang},
  \& {Liu}}]{2016DongVorobyov}
{Dong}, R., {Vorobyov}, E., {Pavlyuchenkov}, Y., {Chiang}, E., \& {Liu}, H.~B.
  2016, \apj, 823, 141

\bibitem[{{Dunham} {et~al.}(2014){Dunham}, {Vorobyov}, \&
  {Arce}}]{2014DunhamVorobyov}
{Dunham}, M.~M., {Vorobyov}, E.~I., \& {Arce}, H.~G. 2014, \mnras, 444, 887

\bibitem[{{Federrath} {et~al.}(2010){Federrath}, {Banerjee}, {Clark}, \&
  {Klessen}}]{2010FederrathBanerjee}
{Federrath}, C., {Banerjee}, R., {Clark}, P.~C., \& {Klessen}, R.~S. 2010,
  \apj, 713, 269

\bibitem[{{Forgan} \& {Rice}(2010)}]{2010ForganRice}
{Forgan}, D. \& {Rice}, K. 2010, \mnras, 402, 1349

\bibitem[{{Green} {et~al.}(2006){Green}, {Hartmann}, {Calvet}, {Watson},
  {Ibrahimov}, {Furlan}, {Sargent}, \& {Forrest}}]{2006ApJ...648.1099G}
{Green}, J.~D., {Hartmann}, L., {Calvet}, N., {et~al.} 2006, \apj, 648, 1099

\bibitem[{{Guo} {et~al.}(2020){Guo}, {Lucas}, {Contreras Pe{\~n}a}, {Kurtev},
  {Smith}, {Borissova}, {Alonso-Garc{\'\i}a}, {Minniti}, {Caratti o Garatti},
  \& {Froebrich}}]{2020ZhenVISTA}
{Guo}, Z., {Lucas}, P.~W., {Contreras Pe{\~n}a}, C., {et~al.} 2020, \mnras,
  492, 294

\bibitem[{{Hackstein} {et~al.}(2015){Hackstein}, {Haas}, {K{\'o}sp{\'a}l},
  {Hambsch}, {Chini}, {{\'A}brah{\'a}m}, {Mo{\'o}r}, {Pozo Nu{\~n}ez},
  {Ramolla}, {Westhues}, {Kaderhandt}, {Fein}, {Barr Dom{\'\i}nguez}, \&
  {Hodapp}}]{2015A&A...582L..12H}
{Hackstein}, M., {Haas}, M., {K{\'o}sp{\'a}l}, {\'A}., {et~al.} 2015, \aap,
  582, L12

\bibitem[{{Hall} {et~al.}(2020){Hall}, {Dong}, {Teague}, {Terry}, {Pinte},
  {Paneque-Carre{\~n}o}, {Veronesi}, {Alexand er}, \& {Lodato}}]{2020Hall}
{Hall}, C., {Dong}, R., {Teague}, R., {et~al.} 2020, arXiv e-prints,
  arXiv:2007.15686

\bibitem[{{Hartmann}(1998)}]{1998Hartmann}
{Hartmann}, L. 1998, {Accretion Processes in Star Formation}

\bibitem[{{Hartmann} \& {Kenyon}(1996)}]{1996HartmannKenyon}
{Hartmann}, L. \& {Kenyon}, S.~J. 1996, \araa, 34, 207

\bibitem[{{Herbig}(1977)}]{1977ApJ...217..693H}
{Herbig}, G.~H. 1977, \apj, 217, 693

\bibitem[{{Hillenbrand} {et~al.}(2018){Hillenbrand}, {Contreras Pe{\~n}a},
  {Morrell}, {Naylor}, {Kuhn}, {Cutri}, {Rebull}, {Hodgkin}, {Froebrich}, \&
  {Mainzer}}]{2018ApJ...869..146H}
{Hillenbrand}, L.~A., {Contreras Pe{\~n}a}, C., {Morrell}, S., {et~al.} 2018,
  \apj, 869, 146

\bibitem[{{Kadam} {et~al.}(2019){Kadam}, {Vorobyov}, {Reg{\'a}ly},
  {K{\'o}sp{\'a}l}, \& {{\'A}brah{\'a}m}}]{2019Kadam}
{Kadam}, K., {Vorobyov}, E., {Reg{\'a}ly}, Z., {K{\'o}sp{\'a}l}, {\'A}., \&
  {{\'A}brah{\'a}m}, P. 2019, \apj, 882, 96

\bibitem[{{Kadam} {et~al.}(2020){Kadam}, {Vorobyov}, {Reg{\'a}ly},
  {K{\'o}sp{\'a}l}, \& {{\'A}brah{\'a}m}}]{2020Kadam}
{Kadam}, K., {Vorobyov}, E., {Reg{\'a}ly}, Z., {K{\'o}sp{\'a}l}, {\'A}., \&
  {{\'A}brah{\'a}m}, P. 2020, \apj, 895, 41

\bibitem[{{Kley} \& {Nelson}(2012)}]{2012KleyNelson}
{Kley}, W. \& {Nelson}, R.~P. 2012, \araa, 50, 211

\bibitem[{{K{\'o}sp{\'a}l} {et~al.}(2016){K{\'o}sp{\'a}l}, {{\'A}brah{\'a}m},
  {Acosta-Pulido}, {Dunham}, {Garc{\'\i}a-{\'A}lvarez}, {Hogerheijde}, {Kun},
  {Mo{\'o}r}, {Farkas}, {Hajdu}, {Hodos{\'a}n}, {Kov{\'a}cs}, {Kriskovics},
  {Marton}, {Moln{\'a}r}, {P{\'a}l}, {S{\'a}rneczky}, {S{\'o}dor},
  {Szak{\'a}ts}, {Szalai}, {Szegedi-Elek}, {Szing}, {T{\'o}th}, {Vida}, \&
  {Vink{\'o}}}]{2016A&A...596A..52K}
{K{\'o}sp{\'a}l}, {\'A}., {{\'A}brah{\'a}m}, P., {Acosta-Pulido}, J.~A.,
  {et~al.} 2016, \aap, 596, A52

\bibitem[{{Kuffmeier} {et~al.}(2018){Kuffmeier}, {Frimann}, {Jensen}, \&
  {Haugb{\o}lle}}]{2018KuffmeierFrimann}
{Kuffmeier}, M., {Frimann}, S., {Jensen}, S.~S., \& {Haugb{\o}lle}, T. 2018,
  \mnras, 475, 2642

\bibitem[{{Lada} {et~al.}(2009){Lada}, {Lombardi}, \&
  {Alves}}]{2009ApJ...703...52L}
{Lada}, C.~J., {Lombardi}, M., \& {Alves}, J.~F. 2009, \apj, 703, 52

\bibitem[{{Lodato} \& {Clarke}(2004)}]{2004LodatoClarke}
{Lodato}, G. \& {Clarke}, C.~J. 2004, \mnras, 353, 841

\bibitem[{{Loinard} {et~al.}(2007){Loinard}, {Torres}, {Mioduszewski},
  {Rodr{\'\i}guez}, {Gonz{\'a}lez-L{\'o}pezlira}, {Lachaume}, {V{\'a}zquez}, \&
  {Gonz{\'a}lez}}]{2007ApJ...671..546L}
{Loinard}, L., {Torres}, R.~M., {Mioduszewski}, A.~J., {et~al.} 2007, \apj,
  671, 546

\bibitem[{{MacFarlane} {et~al.}(2019){MacFarlane}, {Stamatellos}, {Johnstone},
  {Herczeg}, {Baek}, {Chen}, {Kang}, \& {Lee}}]{2019MacFarlane}
{MacFarlane}, B., {Stamatellos}, D., {Johnstone}, D., {et~al.} 2019, \mnras,
  487, 5106

\bibitem[{{Machida} {et~al.}(2011){Machida}, {Inutsuka}, \&
  {Matsumoto}}]{2011MachidaInutsuka}
{Machida}, M.~N., {Inutsuka}, S.-i., \& {Matsumoto}, T. 2011, \apj, 729, 42

\bibitem[{{Magakian} {et~al.}(2013){Magakian}, {Nikogossian}, {Movsessian},
  {Moiseev}, {Aspin}, {Davis}, {Pyo}, {Khanzadyan}, {Froebrich}, {Smith},
  {Moriarty-Schieven}, \& {Beck}}]{2013MNRAS.432.2685M}
{Magakian}, T.~Y., {Nikogossian}, E.~H., {Movsessian}, T., {et~al.} 2013,
  \mnras, 432, 2685

\bibitem[{{Meyer} {et~al.}(2019){Meyer}, {Kreplin}, {Kraus}, {Vorobyov},
  {Haemmerle}, \& {Eisl{\"o}ffel}}]{2019MeyerVorobyov}
{Meyer}, D.~M.~A., {Kreplin}, A., {Kraus}, S., {et~al.} 2019, \mnras, 487, 4473

\bibitem[{{Meyer} {et~al.}(2017){Meyer}, {Vorobyov}, {Kuiper}, \&
  {Kley}}]{2017MeyerVorobyov}
{Meyer}, D.~M.-A., {Vorobyov}, E.~I., {Kuiper}, R., \& {Kley}, W. 2017, \mnras,
  464, L90

\bibitem[{{Movsessian} {et~al.}(2003){Movsessian}, {Khanzadyan}, {Magakian},
  {Smith}, \& {Nikogosian}}]{2003A&A...412..147M}
{Movsessian}, T., {Khanzadyan}, T., {Magakian}, T., {Smith}, M.~D., \&
  {Nikogosian}, E. 2003, \aap, 412, 147

\bibitem[{{Nayakshin} \& {Lodato}(2012)}]{2012NayakshinLodato}
{Nayakshin}, S. \& {Lodato}, G. 2012, \mnras, 426, 70

\bibitem[{{Ninan} {et~al.}(2013){Ninan}, {Ojha}, {Bhatt}, {Ghosh}, {Mohan},
  {Mallick}, {Tamura}, \& {Henning}}]{2013ApJ...778..116N}
{Ninan}, J.~P., {Ojha}, D.~K., {Bhatt}, B.~C., {et~al.} 2013, \apj, 778, 116

\bibitem[{{Peneva} {et~al.}(2010){Peneva}, {Semkov}, \& {Stavrev}}]{Peneva}
{Peneva}, S.~P., {Semkov}, E.~H., \& {Stavrev}, K.~Y. 2010, Bulgarian
  Astronomical Journal, 14, 79

\bibitem[{{Pfalzner}(2008)}]{2008Pfalzner}
{Pfalzner}, S. 2008, \aap, 492, 735

\bibitem[{{Pinte} {et~al.}(2019){Pinte}, {van der Plas}, {M{\'e}nard}, {Price},
  {Christiaens}, {Hill}, {Mentiplay}, {Ginski}, {Choquet}, {Boehler},
  {Duch{\^e}ne}, {Perez}, \& {Casassus}}]{2019Pinte}
{Pinte}, C., {van der Plas}, G., {M{\'e}nard}, F., {et~al.} 2019, Nature
  Astronomy, 3, 1109

\bibitem[{{Reid} {et~al.}(2014){Reid}, {Menten}, {Brunthaler}, {Zheng}, {Dame},
  {Xu}, {Wu}, {Zhang}, {Sanna}, {Sato}, {Hachisuka}, {Choi}, {Immer},
  {Moscadelli}, {Rygl}, \& {Bartkiewicz}}]{2014ApJ...783..130R}
{Reid}, M.~J., {Menten}, K.~M., {Brunthaler}, A., {et~al.} 2014, \apj, 783, 130

\bibitem[{{Reipurth} {et~al.}(2007){Reipurth}, {Aspin}, {Beck}, {Brogan},
  {Connelley}, \& {Herbig}}]{2007AJ....133.1000R}
{Reipurth}, B., {Aspin}, C., {Beck}, T., {et~al.} 2007, \aj, 133, 1000

\bibitem[{{Reipurth} {et~al.}(2012){Reipurth}, {Aspin}, \&
  {Herbig}}]{2012ApJ...748L...5R}
{Reipurth}, B., {Aspin}, C., \& {Herbig}, G.~H. 2012, \apjl, 748, L5

\bibitem[{{Semenov} {et~al.}(2003){Semenov}, {Henning}, {Helling}, {Ilgner}, \&
  {Sedlmayr}}]{2003SemenovHenning}
{Semenov}, D., {Henning}, T., {Helling}, C., {Ilgner}, M., \& {Sedlmayr}, E.
  2003, \aap, 410, 611

\bibitem[{{Semkov} {et~al.}(2019){Semkov}, {Peneva}, {Ibryamov}, \&
  {Mutafov}}]{Semkov}
{Semkov}, E., {Peneva}, S., {Ibryamov}, S., \& {Mutafov}, A. 2019, STARRY Final
  Conference

\bibitem[{{Semkov} {et~al.}(2010){Semkov}, {Peneva}, {Munari}, {Milani}, \&
  {Valisa}}]{2010A&A...523L...3S}
{Semkov}, E.~H., {Peneva}, S.~P., {Munari}, U., {Milani}, A., \& {Valisa}, P.
  2010, \aap, 523, L3

\bibitem[{{Shakura} \& {Sunyaev}(1973)}]{1973ShakuraSunyaev}
{Shakura}, N.~I. \& {Sunyaev}, R.~A. 1973, \aap, 24, 337

\bibitem[{{Staude} \& {Neckel}(1991)}]{1991A&A...244L..13S}
{Staude}, H.~J. \& {Neckel}, T. 1991, \aap, 244, L13

\bibitem[{{Thies} {et~al.}(2010){Thies}, {Kroupa}, {Goodwin}, {Stamatellos}, \&
  {Whitworth}}]{2010Thies}
{Thies}, I., {Kroupa}, P., {Goodwin}, S.~P., {Stamatellos}, D., \& {Whitworth},
  A.~P. 2010, \apj, 717, 577

\bibitem[{{Visser} {et~al.}(2009){Visser}, {van Dishoeck}, {Doty}, \&
  {Dullemond}}]{2009Visser}
{Visser}, R., {van Dishoeck}, E.~F., {Doty}, S.~D., \& {Dullemond}, C.~P. 2009,
  \aap, 495, 881

\bibitem[{{Vorobyov}(2010)}]{2010Vorobyov}
{Vorobyov}, E.~I. 2010, \apj, 713, 1059

\bibitem[{{Vorobyov} {et~al.}(2018){Vorobyov}, {Akimkin}, {Stoyanovskaya},
  {Pavlyuchenkov}, \& {Liu}}]{2018VorobyovAkimkin}
{Vorobyov}, E.~I., {Akimkin}, V., {Stoyanovskaya}, O., {Pavlyuchenkov}, Y., \&
  {Liu}, H.~B. 2018, \aap, 614, A98

\bibitem[{{Vorobyov} \& {Basu}(2005)}]{2005VorobyovBasu}
{Vorobyov}, E.~I. \& {Basu}, S. 2005, \apjl, 633, L137

\bibitem[{{Vorobyov} \& {Basu}(2010)}]{2010VorobyovBasu}
{Vorobyov}, E.~I. \& {Basu}, S. 2010, ApJ, 719, 1896

\bibitem[{{Vorobyov} \& {Basu}(2015)}]{2015VorobyovBasu}
{Vorobyov}, E.~I. \& {Basu}, S. 2015, \apj, 805, 115

\bibitem[{{Vorobyov} \& {Elbakyan}(2018)}]{2018VorobyovElbakyan}
{Vorobyov}, E.~I. \& {Elbakyan}, V.~G. 2018, \aap, 618, A7

\bibitem[{{Vorobyov} {et~al.}(2020{\natexlab{a}}){Vorobyov}, {Elbakyan},
  {Takami}, \& {Liu}}]{2020VorobyovElbakyanTakami}
{Vorobyov}, E.~I., {Elbakyan}, V.~G., {Takami}, M., \& {Liu}, H.~B.
  2020{\natexlab{a}}, \aap, 643, A13

\bibitem[{{Vorobyov} {et~al.}(2020{\natexlab{b}}){Vorobyov}, {Khaibrakhmanov},
  {Basu}, \& {Audard}}]{2020VorobyovKhaibrakhmanov}
{Vorobyov}, E.~I., {Khaibrakhmanov}, S., {Basu}, S., \& {Audard}, M.
  2020{\natexlab{b}}, \aap, 644, A74

\bibitem[{{Vorobyov} {et~al.}(2020{\natexlab{c}}){Vorobyov}, {Skliarevskii},
  {Elbakyan}, {Takami}, {Liu}, {Liu}, \& {Akiyama}}]{2020VorobyovTails}
{Vorobyov}, E.~I., {Skliarevskii}, A.~M., {Elbakyan}, V.~G., {et~al.}
  2020{\natexlab{c}}, \aap, 635, A196

\bibitem[{{Vorobyov} {et~al.}(2017){Vorobyov}, {Steinrueck}, {Elbakyan}, \&
  {Guedel}}]{2017VorobyovSteinrueck}
{Vorobyov}, E.~I., {Steinrueck}, M.~E., {Elbakyan}, V., \& {Guedel}, M. 2017,
  \aap, 608, A107

\bibitem[{{Yang} {et~al.}(2018){Yang}, {Mac Low}, \& {Johansen}}]{2018Yang}
{Yang}, C.-C., {Mac Low}, M.-M., \& {Johansen}, A. 2018, \apj, 868, 27

\bibitem[{{Yorke} \& {Bodenheimer}(2008)}]{2008YorkeBodenheimer}
{Yorke}, H.~W. \& {Bodenheimer}, P. 2008, in Astronomical Society of the
  Pacific Conference Series, Vol. 387, Massive Star Formation: Observations
  Confront Theory, ed. H.~{Beuther}, H.~{Linz}, \& T.~{Henning}, 189

\bibitem[{{Zhu} {et~al.}(2009){Zhu}, {Hartmann}, {Gammie}, \&
  {McKinney}}]{2009ZhuHartmannGammie}
{Zhu}, Z., {Hartmann}, L., {Gammie}, C., \& {McKinney}, J.~C. 2009, \apj, 701,
  620

\bibitem[{{Zhu} {et~al.}(2020){Zhu}, {Jiang}, \& {Stone}}]{2020Zhu}
{Zhu}, Z., {Jiang}, Y.-F., \& {Stone}, J.~M. 2020, \mnras, 495, 3494

\end{thebibliography}

\end{document}